\documentclass[twocolumn,english,prb,amsmath,amssymb,eqsecnum,figsecnum,preprintnumbers]{revtex4}
\usepackage[T1]{fontenc}
\usepackage[latin9]{inputenc}
\usepackage{amstext}
\usepackage{amsmath}
\usepackage{amsfonts}
\usepackage{amssymb}
\usepackage{mathtools}
\usepackage{graphicx}
\usepackage{rotating}
\usepackage {bm}
\usepackage{bbm}
\usepackage[section]{placeins}
\usepackage{color}
\usepackage{multirow}
\usepackage{cellspace} 
\usepackage{enumitem}
\usepackage[scr=rsfs]{mathalfa}
\usepackage{bbold}
\usepackage{array}
\usepackage{multirow}
\usepackage{boldline}
\usepackage[normalem]{ulem}
\makeatletter
\@ifundefined{textcolor}{}
{%
 \definecolor{BLACK}{gray}{0}
 \definecolor{WHITE}{gray}{1}
 \definecolor{RED}{rgb}{1,0,0}
 \definecolor{GREEN}{rgb}{0,1,0}
 \definecolor{BLUE}{rgb}{0,0,1}
 \definecolor{CYAN}{cmyk}{1,0,0,0}
 \definecolor{MAGENTA}{cmyk}{0,1,0,0}
 \definecolor{YELLOW}{cmyk}{0,0,1,0}
}

\def\kF{k_{\text{F}}}
\def\vF{v_{\text{F}}}
\def\TF{T_{\text{F}}}
\def\NF{N_{\text{F}}}

\def\chis{\chi_{\text{s}}}

\def\sgn{{\text{sgn\,}}}

\def\be{\begin{equation}}
\def\ee{\end{equation}}
\def\bea{\begin{eqnarray}}
\def\eea{\end{eqnarray}}
\def\bse{\begin{subequations}}
\def\ese{\end{subequations}}

\def\vso{v_{\text{so}}}

\usepackage{dcolumn}
\usepackage{bm}
\usepackage{amsfonts}
\usepackage{babel}
\draft

\makeatother

\begin{document}
\preprint{arXiv:2602.23554}

\bibliographystyle{unsrtnat}

\title{Generic Long-Range Order-Parameter Correlations in Metallic Quantum Magnets}

\author{T. R. Kirkpatrick$^{1}$ and D. Belitz$^{2,3}$}

\affiliation{$^{1}$ Institute for Physical Science and Technology, University of Maryland, College Park, MD 20742, USA\\
                 $^{2}$ Department of Physics and Institute for Fundamental Science, University of Oregon, Eugene, OR 97403, USA\\
                 $^{3}$ Materials Science Institute, University of Oregon, Eugene, OR 97403, USA\\
 }

\date{\today}
\begin{abstract}
It is shown that in all types of metallic magnets the coupling of the order parameter to the
conduction electrons leads to an order-parameter susceptibility that is long-ranged at zero temperature.
This is true for all known classes of ferromagnets, and also for antiferromagnets and spin-density wave systems, 
helimagnets, magnetic nematics, and altermagnets. The consequences for the magnetic quantum phase
transition vary between different classes of magnets. In almost all 3-d systems with a homogeneous 
magnetization, as well as in magnetic nematics and in altermagnets, the long-ranged correlations
generically modify the nature of the magnetic quantum phase transition from second order to first order. 
The only exception are non-centrosymmetric ferromagnets with a strong spin-orbit interaction, where
the correlations change the order of the transition in 2-d systems, but not in 3-d ones. In helimagnets, 
spin-wave systems, and N{\'e}el antiferromagnets their effect is even weaker and does not 
change the order of the transition if the ordering wave number is sufficiently large, except in flat-band systems. 
In systems with quenched disorder the transition generically is of
second order, but the correlations modify the critical behavior. These conclusions are reached by very 
simple considerations that are based entirely on the single-particle excitations in the nonmagnetic phase 
and their modifications by a field conjugate to the order parameter, augmented by renormalization-group 
considerations.
\end{abstract}
\maketitle

\section{Introduction, and Results}
\label{sec:I}

\subsection{Introduction}
\label{subsec:I.A}

Historically, the two principal types of magnets studied have been ferromagnets and 
antiferromagnets.\cite{Kittel_2004, Blundell_2001} The former, which have been known since antiquity, 
are characterized by a nonzero homogeneous magnetization and, in metallic systems, a spin-split Fermi 
surface. The latter have neither of these features, which makes the magnetic order much 
harder to observe. Both of these magnetic orders, and all others that we will discuss, occur in both
electrically insulating and metallic materials. For our purposes, we will focus on the latter.
Ferrimagnets, i.e., antiferromagnets where the antiparallel magnetic moments do not fully
compensate each other, behave like ferromagnets in this respect. So do spin-canted antiferromagnets; another
class of `imperfect' antiferromagnets where the antiparallel magnetic moments are not collinear, but rather
canted at a small angle. Spin canting can also occur in ferromagnets, where it reduces the homogeneous 
magnetization. More exotic types of magnetic order include helimagnets\cite{Blundell_2001} and magnetic 
nematics.\cite{Wu_et_al_2007} In the former the magnetization forms a global spiral, often with a long wavelength. 
The homogeneous magnetization vanishes, but the spiral order splits the Fermi surface. In the latter, the 
electron-electron interaction leads to a spin-dependent Pomeranchuk instability;\cite{Pomeranchuk_1958} 
this results in a split Fermi surface, but no homogeneous magnetization. 

In recent years a new class of magnets has been proposed, see 
Refs.~\onlinecite{Hayami_Yanagi_Kusunosel_2019, Smejkal_et_al_2020, Yuan_et_al_2020, Mazin_et_al_2021, Ma_et_al_2021, Hu_et_al_2025}, 
a subclass of which are now known as `altermagnets'.\cite{Smejkal_Sinova_Jungwirth_2022, Yuan_Georgescu_Rondinelli_2024, Song_et_al_2025} 
They combine antiferromagnetic order with a split distorted Fermi surface reminiscent of magnetic nematics. However, 
while the split Fermi surface in magnetic nematics arises from electron-electron interaction effects, in 
altermagnets it results from specific features of the band structure. Many altermagnetic candidates are not 
metallic; however, the relevant band structure has been predicted and observed in the semiconductor 
MnTe,\cite{Mazin_2023, Krempasky_et_al_2024} and time-reversal symmetry breaking consistent with 
altermagnetism has been observed in the metallic compound RuO$_2$.\cite{Fedchenko_et_al_2024} 
Another metallic system that has been predicted to harbor altermagnetism is doped 
FeSb$_2$.\cite{Mazin_et_al_2021} In addition to being of fundamental interest, these systems
are interesting for spintronics applications.\cite{Duan_et_al_2025, Gu_et_al_2025, Smejkal_2024}

An interesting question is the nature of the quantum phase transition that can be probed by tuning some
non-thermal control parameter, often pressure, to drive the system from the magnetically disordered to the
magnetically ordered phase at a fixed low temperature (zero in idealized theoretical treatments). Since the
quantum fluctuations at zero temperature are very different from the thermal ones that drive the phase
transition at nonzero temperature, and since in quantum systems the statics and the dynamics are 
coupled, the quantum critical behavior is expected to be different from the
critical behavior at the thermal transition. For metallic ferromagnets, this problem was first investigated by
Hertz,\cite{Hertz_1976} who concluded that the quantum critical behavior is generically mean-field like
with the statics described by a simple Landau theory and the dynamics by van Hove theory. 
The basic argument is that in quantum statistical mechanics 
the coupling between the statics and the dynamics leads to an effective system dimensionality 
$d_{\text{eff}} = d+z$, with $d$ the spatial dimensionality and $z$ the dynamical critical exponent, 
that exceeds the upper critical dimensionality $d_{\text{uc}} = 4$ above which fluctuations are irrelevant for the 
leading critical behavior. This conclusion turned out to be in general not correct. The reason is the existence, 
in most types of metals, of soft or massless electronic excitations that are rendered massive by the magnetic 
order parameter. As a result, the mean-field free energy is not an analytic function of the order parameter and 
the simple Landau theory breaks down. This is well established for metallic ferromagnets,\cite{Belitz_Kirkpatrick_Vojta_1999} 
where a generalized mean-field theory that takes the nonanalyticies into account is in excellent agreement
with experiment and sometimes referred to as BKV theory;\cite{Huang_et_al_2016, Mishra_et_al_2020}
for a review, see Ref.~\onlinecite{Brando_et_al_2016a}. It is also true in magnetic nematics\cite{Kirkpatrick_Belitz_2011} 
and, as we will show, altermagnets. In this paper we further show that, remarkably, it also is
true in helimagnets and in N{\'e}el antiferromagnets.\cite{2d_AFM_footnote} 
The strength of the effect varies between classes of magnets;
it is stronger in classes with a vanishing ordering wave vector and weaker in helimagnets and
antiferromagnets. It also is qualitatively different in systems with quenched disorder compared to
clean ones. As a result, the nonanalyticity in the free energy appears at different orders in the
Landau expansion for different classes of systems and modifies the nature of the transition 
described by ordinary Landau theory only for certain classes. 

Underlying the breakdown of the Landau theory are long-ranged correlations of the order parameter
fluctuations that result from the coupling of the latter to fermionic soft modes. The main purpose of
the present paper is to show that these correlations
are present in all metallic magnets; the only question is whether their effect is strong enough to
modify the mean-field critical behavior predicted by ordinary Landau theory. Importantly, they
are present even in the nonmagnetic phase. In the second part of this section we show
how one can deduce these correlations and their impact on the quantum phase transition from 
very simple scaling arguments. In the remainder of the paper we give a comprehensive discussion 
of these phenomena for all known classes of magnets and provide a very simple criterion for whether 
or not long-ranged correlations are present. In Sec.~\ref{sec:II} we introduce the relevant fermionic 
soft modes and provide the basic criterion for whether they lead to a breakdown of the Landau theory 
for the magnetic order. In Sec.~\ref{sec:III} we apply this criterion to the various classes of magnets, and
in Sec.~\ref{sec:IV} we provide a summary and conclusion. Some technical aspects regarding
helimagnets and antiferromagnets are relegated to two appendices.

\subsection{Simple Arguments, and Results}
\label{subsec:I.B}

Let $\mathfrak{O}$ be an observable, $\mathfrak{h}$ the field conjugate to $\mathfrak{O}$, and
$\chi_{\mathfrak{O}}$ the corresponding susceptibility. For our purposes, $\mathfrak{O}$ is a
magnetization that in general is modulated by an ordering wave vector ${\bm Q}$ with modulus
$\vert{\bm Q}\vert = Q$. Its average amplitude serves as the order parameter 
$\Phi$ for the magnetic phase transition related to ${\mathfrak O}$. If the magnetization is modulated, 
then so is the conjugate field $\mathfrak{h}$.

\subsubsection{Scaling of the order-parameter susceptibility}
\label{subsubsec:I.B.1}

We start by analyzing the behavior of $\chi_{\mathfrak O}$ by means of standard scaling arguments.\cite{Ma_1976}
If the fermion system contains soft modes that couple to ${\mathfrak O}$ (see Sec.~\ref{sec:II} 
for the relevant criterion), then we expect $\chi_{\mathfrak{O}}$ to contain a part $\delta\chi_{\mathfrak{O}}$ 
that obeys a scaling law
\be
\delta\chi_{\mathfrak{O}}(q,\mathfrak{h}) = b^{-(d-1)} S_{\mathfrak{O}}(q b,\mathfrak{h} b)
\label{eq:1.1}
\ee
with $S_{\mathfrak O}$ a scaling function.
Here $q$ is the wave number, and $b>0$ is the (arbitrary) length scaling factor. We have
chosen the wave number to have a scale dimension $[q] = 1$, which is reflected in the first
argument of the scaling function. Accordingly, length $L$ and volume
$V$ have scale dimensions $[L] = -1$ and $[V] = -d$, respectively. In a generic clean fermionic system 
an energy $E$ scales as the wave number, and hence $[E] = 1$. (An exception are flat-band systems,
where $E$ scales as a higher power of the wave number, which has interesting consequences, see
Sec.~\ref{subsubsec:IV.B.4}.) $\mathfrak{h}$ enters via a
Zeeman-like term, and hence its scale dimension is $[\mathfrak{h}] = [E] = 1$, which determines
the second argument of the scaling function. The susceptibility
is the second derivative of the free-energy density with respect to $\mathfrak{h}$, so its scale
dimension is $[\delta\chi_{\mathfrak{O}}] = [1/EV] = d-1$, which is reflected in the prefactor
on the right-hand side of Eq.~(\ref{eq:1.1}). For zero field, Eq.~(\ref{eq:1.1}) yields a nonanalytic 
wave-number dependence
\bse
\label{eqs:1.2}
\be
\chi_{\mathfrak{O}}(q\to 0,\mathfrak{h}=0) \propto q^{d-1}\ .
\label{eq:1.2a}
\ee
For $1<d<3$ this is the leading wave-number dependence, for $d>3$ it is subleading
to an analytic $q^2$ contribution. For $d=3$ one expects a $q^2 \ln q$ behavior that cannot
be captured by scaling considerations alone. In real space this corresponds to a power-law
decay at long distances,\cite{Lighthill_1958}
\be
\chi_{\mathfrak{O}}(r\to\infty,\mathfrak{h}=0) \propto 1/r^{2d-1}\ .
\label{eq:1.2b}
\ee
\ese
This is an example of `generic scale invariance', i.e., power-law behavior that is present in an
entire phase and not tied to special points in the phase diagram.\cite{Belitz_Kirkpatrick_Vojta_2005}
What underlies this generic scale invariance is a stable (as opposed to critical) renormalization-group
fixed point.\cite{Ma_1976, Anderson_1984} In the current context the stable fixed point in question
describes the Fermi-liquid phase.\cite{Belitz_Kirkpatrick_2014}
For zero wave number, there is a corresponding nonanalyticity with respect to the field,
\be
\chi_{\mathfrak{O}}(q = 0,\mathfrak{h}) \propto \mathfrak{h}^{d-1}
\label{eq:1.3}
\ee
and same comments as for Eq.~(\ref{eq:1.2a}) apply. 

If the ordering wave vector ${\bm Q}$ is nonzero, then the above expressions require some
qualifying remarks. First, the wave number $q$ in Eqs.~(\ref{eq:1.1}) and (\ref{eq:1.2a}) is
measured with respect to ${\bm Q}$ rather than with respect to zero wave number. Second,
the nonanalyticity in Eq.~(\ref{eq:1.3}) will have a zero prefactor since the average of
${\mathfrak h}$ vanishes. The leading ${\mathfrak h}$-nonanalyticity must therefore be
given by the corresponding power of ${\mathfrak h}^2$, whose average is nonzero, with a factor of $\vF Q$
(with $\vF$ the fermionic velocity scale) restoring the proper dimensionality. We thus expect
\be
\chi_{\mathfrak{O}}(q = 0,\mathfrak{h}) \propto( \mathfrak{h}^2/\vF Q)^{d-1}
\label{eq:1.4}
\ee
From the general scaling law this can be seen as follows. The additional length scale
$\lambda \sim 1/Q$ requires a modification of the scaling law; Eq.~(\ref{eq:1.1}) now reads
\be
\delta\chi_{\mathfrak{O}}(q,\mathfrak{h}) = b^{-(d-1)} S_{\mathfrak{O}}(q b, \mathfrak{h} b, \lambda b^{-1})\ .
\label{eq:1.5}
\ee
For $q=0$ this yields
\be
\delta\chi_{\mathfrak{O}}(q=0,\mathfrak{h}) = {\mathfrak h}^{d-1} S_{\mathfrak{O}}(0, 1, \lambda \mathfrak{h})\ .
\label{eq:1.6}
\ee
If $S_{\mathfrak{O}}(0,1,0)$ were a constant, this would again yield Eq.~(\ref{eq:1.3}). However, as
we will see in Sec.~\ref{sec:III}, $S_{\mathfrak{O}}(0,1,x\to 0) \propto x^{d-1}$. That is, the microscopic
length scale $\lambda$, which is irrelevant (in the sense of the renormalization group) by power
counting, serves as a dangerous irrelevant variable\cite{Ma_1976, Fisher_1983} and we obtain Eq.~(\ref{eq:1.4}). 

We can combine the cases of zero and nonzero $Q$ by writing 
\be
\delta\chi_{\mathfrak{O}}(q,\mathfrak{h}) = b^{-(d-1)} S_{\mathfrak{O}}(q b,\mathfrak{h} b^{[{\mathfrak h}]})
\ee
where the scale dimension of the field is $[{\mathfrak h}] = 1$ for $Q=0$ and, effectively, $[{\mathfrak h}]=1/2$
for $Q \neq 0$. 

In the presence of quenched disorder the electron dynamics are diffusive and hence the energy 
scales as the wave number squared, i.e., its scale dimension is equal to $2$. Accordingly, for 
magnets with a zero ordering wave number Eq.~(\ref{eq:1.1}) gets modified to
\be
\delta\chi_{\mathfrak{O}}(q,\mathfrak{h}) = b^{-(d-2)} S_{\mathfrak{O}}(q b, \mathfrak{h} b^2)\ .
\label{eq:1.8}
\ee
In zero field we now have
\bse
\label{eqs:1.9}
\bea
\delta\chi_{\mathfrak{O}}(q\to 0) &\propto& q^{d-2} 
\label{eq:1.9a}\\
\delta\chi_{\mathfrak{O}}(r\to \infty) &\propto& 1/r^{2(d-1)} \ ,
\label{eq:1.9b}
\eea
\ese
and at zero wave number,
\be
\delta\chi_{\mathfrak{O}}(\mathfrak{h}\to 0) \propto \mathfrak{h}^{(d-2)/2}\ .
\label{eq:1.10}
\ee
A nonzero ordering wave number acts as a dangerous irrelevant variable with respect to the field
scaling as in the clean case, and we obtain
\be
\delta\chi_{\mathfrak{O}}(q=0,\mathfrak{h})  \propto (\mathfrak{h}^2/\vF Q)^{(d-2)/2}\ .
\label{eq:1.11}
\ee
The above expressions are valid for $2<d<4$. For $d=2$ the description breaks down because of
electron localization effects. 

The sign of the nonanalytic field dependence is important in what follows and is determined by 
general physical arguments. They were given in Refs.~\onlinecite{Belitz_Kirkpatrick_2014} and
\onlinecite{Brando_et_al_2016a} for the case of ferromagnetic order and generalize to any kind 
of magnetic order as follows. In clean systems, the fluctuations that are responsible for the 
nonanalyticity decrease the tendency of the system to order magnetically, and therefore the fluctuation
contribution to the zero-field susceptibility is negative. A nonzero field weakens the fluctuations,
and hence the leading effect of the field increases $\chi_{\mathfrak{O}}$ from its zero-field value.
The prefactor in Eqs.~(\ref{eq:1.3}) and (\ref{eq:1.4}) is therefore positive. In disordered metals,
the slow (diffusive) dynamics of the electrons increase the tendency toward magnetic order. A
nonzero field suppresses the diffusive fluctuations and hence decreases the susceptibility. 
Accordingly, the prefactor of the nonanalyticity in Eqs.~(\ref{eq:1.10}) and (\ref{eq:1.11}) is negative.

\subsubsection{The nonanalytic free-energy functional}
\label{subsubsec:I.B.2}

Since the susceptibility is the second derivative of the free energy with respect to the conjugate field, 
a nonanalyticity in $\chi_{\mathfrak O}$ implies a nonanalyticity in a generalized mean-field free energy
that ignores fluctuations of the order parameter $\Phi$. The scaling of the free energy with respect to 
$\Phi$ is the same as the scaling with respect to the conjugate field, and therefore Eqs.~(\ref{eq:1.3}), 
(\ref{eq:1.4}), (\ref{eq:1.10}), and (\ref{eq:1.11}) imply specific nonanalyticities of the free-energy 
functional $f$. In clean systems we have
\bse
\label{eqs:1.12}
\be
f(\Phi) = -\mathfrak{h}\Phi + r\,\Phi^2 - v\,\Phi^{d+1} + u\,\Phi^4 + O(\Phi^6)
\label{eq:1.12a}
\ee
if the ordering wave number $Q$ is zero, and
\be
f(\Phi) = -\mathfrak{h}\Phi + r\,\Phi^2 + u\,\Phi^4 + O(\Phi^{2d})
\label{eq:1.12b}
\ee
if $Q \neq 0$. 
\ese
Here $r$, $v$, and $u$ are Landau coefficients. Importantly, $v>0$ by the arguments given after Eq.~(\ref{eq:1.11}). 
In $d=3$ the nonanalytic terms
in Eqs.~(\ref{eq:1.12a}) and (\ref{eq:1.12b}) should be interpreted as $\Phi^4 \ln(1/\Phi)$ and 
$\Phi^6 \ln(1/\Phi)$, respectively. In 2-d, and for $Q\neq 0$, the soft-mode contribution competes
with the $\Phi^4$ term in the ordinary Landau theory, but its prefactor has a definite (negative)
sign. 

In the absence of fermionic soft modes, $v=0$ and $f$ is an ordinary Landau free energy that generically
(i.e., for $u>0$) describes a second-order transition with mean-field exponents. For $1<d<3$
the soft modes change this to a first-order transition if $Q=0$. This is true for all known classes of
magnets with a vanishing ordering wave vector. It remains true for $d=3$ with the exception of 
non-centrosymmetric ferromagnets with a strong spin-orbit coupling, see Sec.~\ref{subsubsec:III.A.4}.
For a nonzero ordering wave vector the free-energy functional is still nonanalytic, but the nonanalyticity
occurs at too high an order in the Landau expansion to change the order of the transition, see
Eq.~(\ref{eq:1.12b}). Note that this is entirely a result of the scale dimension of the conjugate field,
which is different in the two cases; the scaling with the wave number, and the resulting long-rangedness
of order-parameter correlations in real space, is the same whether or not ${\bm Q} = 0$. In 2-d
systems with $Q\neq 0$ the soft-mode contribution competes with the quartic term in the
ordinary Landau theory and may or may not lead to a first-order transition.

In systems with quenched disorder the nonanalyticity is stronger than in clean ones, and the sign
changes. We have
\bse
\label{eqs:1.13}
\be
f(\Phi) = -{\mathfrak h} \Phi + r \Phi^2 + v\Phi^{(d+2)/2} + u \Phi^4 + O(\Phi^6)
\label{eq:1.13a}
\ee
if $Q=0$, and
\be
f(\Phi) = -{\mathfrak h} \Phi + r \Phi^2 + v \Phi^d + u \Phi^4  + O(\Phi^6)
\label{eq:1.13b}
\ee
\ese
if $Q \neq 0$. Again, $v>0$. In this case, the transition is still of second order, but the nonanalyticity
changes the critical behavior for all dimensions $2<d<6$ if $Q=0$, and for $2<d<4$ if $Q \neq 0$. 

The effects of order-parameter fluctuations depend strongly on the class of magnet and the nature
of the generalized mean-field theory. If the latter predicts a first-order transition, then their effects are
expected to be weak. This is due to the central argument of Hertz theory: In the absence of the 
long-range correlations that lead to the first-order transition the system would be above its upper
critical dimension and therefore order-parameter fluctuations are unimportant. In the case of
disordered ferromagnets, where the generalized mean-field theory is given by Eq.~(\ref{eq:1.13a}),
they lead to complicated logarithmic corrections to scaling, see the discussion in Sec.~\ref{subsubsec:III.A.1}.
For antiferromagnets, the quantum criticality is a very complicated problem.\cite{AFM_QCP_footnote}
For helimagnets, the effects of order-parameter fluctuations are not known.

\section{A Classification of Metallic Magnets}
\label{sec:II}

\subsection{Soft modes in metals}
\label{subsec:II.A}

In this section we identify the soft modes that underly the scaling behavior in Sec.~\ref{subsec:I.B}.

\subsubsection{A simple example}
\label{subsubsec:II.A.1}

Soft, or massless, modes or excitations are correlation functions that diverge as the external frequency
and wave number go to zero. In metals, a  class of soft modes that are relevant in many
different contexts is given by convolutions of single-particle Green functions. To make a point we consider,
as a very simple example, noninteracting conduction electrons with a single-particle energy $\epsilon_{\bm k}$
and chemical potential $\mu$. The single-particle Hamiltonian is diagonal in wave-vector space, and the
diagonal matrix elements are
\bse
\label{eqs:2.1}
\be
{\mathcal H}_{\bm k} = \xi_{\bm k}\,\sigma_0 - {\bm h}\cdot{\bm\sigma}\ .
\label{eq:2.1a}
\ee
Here $\xi_{\bm k} = \epsilon_{\bm k} - \mu$, $\bm\sigma = (\sigma_1,\sigma_2,\sigma_3)$ are the Pauli
matrices, $\sigma_0$ is the $2\times 2$ unit matrix, and ${\bm h}$ is an external magnetic field. 
For simplicity we will assume a quadratic and isotropic single-electron energy 
\be
\epsilon_{\bm k} = {\bm k}^2/2m
\label{eq:2.1b}
\ee
\ese
throughout most of this paper.\cite{single-electron-spectrum_footnote} The single-particle Green function is defined as
\be
{\mathcal G}_k = \left[i\omega_m\,\sigma_0 - {\mathcal H}_{\bm k}\right]^{-1}\ ,
\label{eq:2.2}
\ee
where $k = (i\omega_m,{\bm k})$ is a 4-vector that comprises a fermionic Matsubara frequency $\omega_m$
and a wave vector ${\bm k}$. The Green function can be expressed as a superposition of quasiparticle
resonances
\be
F_k^{\sigma} = \frac{1}{i\omega_m - \xi_{\bm k} - \sigma h}
\label{eq:2.3}
\ee
as follows:
\bse
\label{eqs:2.4}
\be
{\mathcal G}_k = \sum_{\sigma = \pm} F_k^{\sigma} {\mathcal M}^{\sigma}(\hat{\bm h})\ .
\label{eq:2.4a}
\ee
Here $h = \vert{\bm h}\vert$, $\hat{\bm h} = {\bm h}/h$ and
\be
{\mathcal M}^{\sigma}(\hat{\bm h}) = \frac{1}{2} \left(\sigma_0 - \sigma \hat{\bm h}\cdot{\bm\sigma}\right)\ .
\label{eq:2.4b}
\ee
\ese
Note that ${\mathcal H}_{\bm k}$, ${\mathcal G}_k$, and 
${\mathcal M}^{\sigma}$ are $2\times 2$ matrices, whereas the quasiparticle resonance $F_k^{\sigma}$ is scalar
 valued. 

Now consider a wave-vector convolution of two quasiparticle resonances,
\bse
\label{eqs:2.5}
\be
\varphi_{i\omega_m}^{\sigma\sigma'}({\bm q},i\Omega_n) = \frac{1}{V} \sum_{\bm k} F_k^{\sigma} F_{k-q}^{\sigma'}
\label{eq:2.5a}
\ee
where $q \equiv (i\Omega_n,{\bm q})$ comprises a bosonic Matsubara frequency $\Omega_n$ and
a wave vector ${\bm q}$, and $V$ is the system volume. In an approximation that captures the leading behavior for small ${\bm q}$ and 
$\Omega_n$\cite{Abrikosov_Gorkov_Dzyaloshinski_1963} we obtain
\bea
\varphi_{i\omega_m}^{\sigma\sigma'}({\bm q},i\Omega_n) &=& 2\pi i \, \sgn(\omega_m)\, \Theta(-\omega_m(\omega_m - \Omega_n))
\nonumber\\
&& \hskip 50pt \times \phi^{\sigma\sigma'}({\bm q},i\Omega_n)\ ,
\label{eq:2.5b}
\eea
where
\be
\phi^{\sigma\sigma'}({\bm q},i\Omega_n) = \NF
   \int \frac{d\Omega_{\bm k}}{S_{d-1}}\,\frac{1}{i\Omega_n - \vF\hat{\bm k}\cdot{\bm q} - (\sigma - \sigma')h}
\label{eq:2.5c}
\ee
\ese
with $S_{d-1}$ the surface area of the $(d-1)$-sphere. 
Here $\NF$ and $\vF$ are the density of states at the Fermi surface and the Fermi velocity, respectively,
and $\Omega_{\bm k}$ is the solid angle with respect to ${\bm k}$. Here, and throughout the paper, the
Fermi wave number $\kF$, and all quantities derived from it, is to be interpreted as the free-fermion
Fermi wave number in zero field, $\kF = \sqrt{2m\mu}$.

The salient point, for our purposes, is that this correlation function, for $h=0$ and for $\omega_m$ and 
$\omega_m - \Omega_n$ having different signs, diverges as $\Omega_n, {\bm q} \to 0$
and hence is a soft mode. A natural question regards the robustness of this property, as we have demonstrated it
for noninteracting electrons only. The key to the answer lies in the partial fraction decomposition known as Velicky's 
Ward identity,\cite{Velicky_1969}
\bea
F_{i\omega_{m_1},{\bm k}+{\bm q}/2}^{\sigma}\, F_{i\omega_{m_2},{\bm k}-{\bm q}/2}^{\sigma'} &=&
\nonumber\\
&& \hskip -80pt \frac{-\left(F_{i\omega_{m_1},{\bm k}+{\bm q}/2}^{\sigma} - F_{i\omega_{m_2},{\bm k}-{\bm q}/2}^{\sigma'}\right)}
       {i\Omega_{m_1-m_2} - {\bm k}\cdot{\bm q}/m - (\sigma - \sigma')h}
\label{eq:2.6}
\eea
with $\Omega_{m_1-m_2} = \omega_{m_1} - \omega_{m_2}$.        
Equation~(\ref{eq:2.6}) relates a four-fermion correlation function on the left-hand side to the difference between
two two-fermion correlation functions on the right-hand side. Now consider the limit ${\bm q}\to 0$ and 
$\omega_{m_1}, \omega_{m_2} \to 0$. Since $F$ as a function of its complex frequency has a cut along
the real axis, the numerator on the right-hand side is nonzero in this limit if $\omega_{m_1}$ and $\omega_{m_2}$
have different signs, whereas the denominator vanishes if $h=0$ (or if $\sigma = \sigma'$). The soft nature of
the four-fermion correlation is thus tied to the analytic structure of the single-particle Green function, which is
very robust. In particular, the qualitative aspects of the soft mode persist in the presence of an electron-electron
interaction, provided the left-hand side of Eq.~(\ref{eq:2.6}) is replaced by the appropriate four-fermion correlation
function that factorizes into a product of two Green functions in the noninteracting limit.\cite{Belitz_Kirkpatrick_2012a}
It further remains intact in the presence of quenched disorder, despite the fact that the disorder gives the single-particle
Green function a mass.\cite{Belitz_Kirkpatrick_1997} The only difference in this case is that the soft mode becomes
diffusive, i.e., the frequency $\Omega$ scales quadratically with the wave number ${\bm q}$, rather than linearly
as in the ballistic mode in Eq.~(\ref{eq:2.6}). These `diffusons' in disordered systems can be understood as the 
Goldstone modes of a spontaneously broken rotational symmetry in frequency space, i.e., a symmetry between 
retarded and advanced degrees of freedom, as was first shown by Wegner.\cite{Wegner_1979, Schaefer_Wegner_1980} 
This interpretation of the robustness of the soft mode carries over to clean systems.\cite{Belitz_Kirkpatrick_2012a}
We also note that at any nonzero temperature interactions cause the soft mode to acquire a mass whose physical
interpretation is a dephasing rate.\cite{Raimondi_Schwab_Castellani_1999} However, this rate scales with
the temperature as $T^{d-1}$ in $d$ spatial dimensions and in bulk systems is subleading to the $\Omega\sim T$
scaling of the frequency. 

The above considerations capture the essence of the soft modes relevant for this paper in a simple metal
where the conduction electrons form a Landau Fermi liquid. We will see later that analogous considerations
hold in more complicated metals.

\subsubsection{Two kinds of soft modes}
\label{subsubsec:II.A.2}       

An important aspect of the soft modes is that they fall into two different classes.\cite{Belitz_Kirkpatrick_Vojta_2002}
Using the same notation as in Sec.~\ref{subsec:I.B}, let $\mathfrak{O}$ be an observable and let $\mathfrak{h}$ be 
the field conjugate to $\mathfrak{O}$. Let $M$ be a mode that is soft if $\mathfrak{h} = 0$. Using the terminology of 
Ref.~\onlinecite{Kirkpatrick_Belitz_2019a},  we say that $M$, with respect to $\mathfrak{h}$, is

$\bullet$
\hangindent=18pt
\emph{a soft mode of the first kind} (SM1) if $\mathfrak{h}\neq 0$ gives $M$ a mass,

\par\noindent
or

$\bullet$ 
\hangindent=18pt
\emph{a soft mode of the second kind} (SM2) if $M$ remains soft for $\mathfrak{h}\neq 0$.

\noindent
An equivalent statement is that a SM1 results if the conjugate field splits the Fermi surface, or parts
of it, whereas a SM2 results if sheets of the Fermi surface that are degenerate in zero field remain
degenerate in a nonzero field.\cite{FS_splitting_footnote}

This concept can also be illustrated by means of the simple example in Sec.~\ref{subsubsec:II.A.1}:
Take the observable in question to be the spin density. Then the magnetic field ${\bm h}$ is the
relevant conjugate field. Of the soft modes $\varphi^{\sigma\sigma'}$ in Eqs.~(\ref{eqs:2.5}), those
with $\sigma \neq \sigma'$ are soft of the first kind with respect to ${\bm h}$, whereas those
with $\sigma = \sigma'$ are soft of the second kind.        

To see the significance of this distinction, consider the partition function $Z[\mathfrak{h}]$ as a
generating functional of $\mathfrak{O}$-correlation functions:
\be
Z[\mathfrak{h}] = \int D[\bar\psi,\psi]\,e^{S[\bar\psi,\psi;\mathfrak{h}]}
\label{eq:2.7}
\ee
Here $S$ is the action in terms of Grassmann-valued spinor fields $\bar\psi$ and $\psi$.\cite{Negele_Orland_1988}
It consists of three parts,
\bse
\label{eqs:2.8}
\be
S[\bar\psi,\psi;\mathfrak{h}] = S_0 + S_{\mathfrak{h}} + S_{\text{int}}
\label{eq:2.8a}
\ee
namely, the free-electron action in zero field
\be
S_0 = \sum_{{\bm k},i\omega_m}  \bar\psi(i\omega_m,{\bm k}) \left[i\omega_m \sigma_0
     - {\mathcal H}_{{\bm k},\,{\mathfrak{h}=0}}\right] \psi(i\omega_m,{\bm k})
\label{eq:2.8b}
\ee
a field-dependent term of Zeeman type,
\be
\label{eq:2.8c}
S_{\mathfrak{h}} = \mathfrak{h} \int_V d{\bm x} \int_0^{1/T}\!\! d\tau\ \mathfrak{O}({\bm x},\tau)\ ,
\ee
\ese
and a part $S_{\text{int}}$ that describes the electron-electron interaction. 
In Eq.~(\ref{eq:2.8c}) ${\bm x}$ is the real-space position, 
and $\tau$ is the imaginary-time variable. If the system contains any soft modes
of the first kind with respect to $\mathfrak{h}$, then $Z$ cannot be an analytic function of $\mathfrak{h}$.
The same is true for the free energy density $f[\mathfrak{h}] = -(T/V) \ln Z[\mathfrak{h}]$ and all of its derivatives, 
in particular the $\mathfrak{O}$-susceptibility $\chi_{\mathfrak{O}} = \partial^2 f/\partial \mathfrak{h}^2$.
The only caveat is that this conclusion requires the presence of an electron-electron interaction in 
the appropriate channel that leads to a mixing of positive and negative Matsubara frequencies. In
other words, the observable must couple to the soft mode, but generically all couplings allowed
by symmetry will be present in an interacting electron system. 
The conclusion is that modes that are soft of the first kind {\em with respect to a given observable} $\mathfrak{O}$
and its conjugate field $\mathfrak{h}$ lead to nonanalytic behavior of the $\mathfrak{O}$-susceptibility
and to long-range correlations of the $\mathfrak{O}$-fluctuations, see below. This in turn will in general
affect any phase transition for which $\mathfrak{O}$ serves as an order parameter, see Sec.~\ref{subsec:II.B}
below. Soft modes of the second kind have none of these consequences, although they of course still enter
the free energy and may influence the behavior of other observables.

Coming back to the example above, the spin susceptibility $\chi_{\text{s}}$ in an interacting electron system is a 
nonanalytic function of the magnetic field. The leading nonanalytic behavior in a clean system in $d$ spatial dimensions 
is given by
\be
\chis(h\to 0) = \text{const.} + \begin{cases}  a_d\,h^{d-1}    &  \text{for}\quad  1<d<3 \\
                                                          a_3\,h^2 \ln(1/h) & \text{for} \quad d=3
                                                          \end{cases} \ ,
\label{eq:2.9}
\ee
with a positive prefactor $a_d>0$. This is a specific example of the behavior described by Eq.~(\ref{eq:1.3}). 
For $d=3$ this result was first obtained within the framework of Landau
Fermi-liquid theory.\cite{Misawa_1971} It was later understood to be a manifestation of a more general
scaling behavior related to the fact that $\chis$ is a nonanalytic function of the wave
number, and also the temperature,\cite{Belitz_Kirkpatrick_Vojta_1997, Chitov_Millis_2001, 
Galitski_Chubukov_Das_Sarma_2005, Betouras_Efremov_Chubukov_2005}
and renormalization-group techniques have shown that the exponent represents the exact
leading nonanalyticity.\cite{Belitz_Kirkpatrick_2014} The nonanalytic behavior indicates the presence of
long-range correlations and generic scale invariance, and the role of the underlying soft modes has
been stressed in Ref.~\onlinecite{Belitz_Kirkpatrick_Vojta_2005}. In a clean electron system the magnetic field $h$, 
the wave number $q$, and the temperature $T$ all scale the same way, $h\sim q\sim T$, and therefore 
analogous scaling behaviors hold with $h$ replaced by $q$ or $T$, see Eqs.~(\ref{eqs:1.2}).\cite{T_scaling_footnote}
In real space this corresponds to long-range correlations of the form
\be
\chis(r\to\infty) \propto \begin{cases} 1/r^{2d-1} & \text{for} \quad 1<d<3 \\
                                                          \ln r/r^5 & \text{for} \quad d=3
                                                          \end{cases}\ .
\label{eq:2.10}
\ee
The sign of the nonanalyticity, which will be important later, comes about as follows.\cite{Belitz_Kirkpatrick_2014, Brando_et_al_2016a} The fluctuations that are responsible for the nonanalyticity decrease the tendency of the system
to order ferromagnetically, and therefore the fluctuation contribution to the zero-field susceptibility is negative. A 
magnetic field weakens the fluctuations, and hence the leading effect of the field increases $\chis$ from
its zero-field value. We re-iterate that this argument holds much more generally than for the simple ferromagnetic 
example, see Sec.~\ref{subsec:I.B}. As mentioned above, in order for the prefactor $a_d$ to be nonzero the presence of
an appropriate frequency-mixing interaction is required. Since only the modes with $\sigma \neq \sigma'$
are soft of the first kind with respect to $h$, the relevant interaction in this example is the spin-triplet interaction in the
channels given by the Pauli matrices $\sigma_1$ and $\sigma_2$. If these interaction amplitudes are
not included in the bare action they will be generated under renormalization by interactions in any other
channels. In perturbation theory, the relevant frequency mixing first occurs at quadratic order in the
spin-triplet interaction coefficient.

The presence of quenched disorder leads to several important changes. First, the scale dimension
of the coupling constant that leads to the nonanalyticity changes from $d-1$ to $d-2$. This is because
the electron dynamics are now diffusive rather than ballistic. Second, the energy scales
$\Omega$, $T$, and $h$ now scale as the square of the wave number, $\Omega \sim T \sim h \sim q^2$.
Third, the sign of the nonanalytic term changes. This is because in a disordered system fluctuations
slow down the dynamics of the electrons, which enhances the tendency of the system to order
ferromagnetically. Accordingly, fluctuation contribution to the zero-field susceptibility is positive,
and an external field, which weakens the fluctuations, decreases $\chis$ from its zero-field value. 
One therefore has
\be
\chis = \text{const.} - {\tilde a}_d \,h^{(d-2)/2} \qquad \text{for} \quad 1<d\leq 3\ .
\label{eq:2.11}
\ee
with ${\tilde a}_d > 0$. Again this is just a specific example of the more general behavior given
in Eq.~(\ref{eq:1.10}). This specific result also was first derived within perturbation
theory\cite{Altshuler_Aronov_Zyuzin_1983} and later corroborated and shown to be exact by means 
of renormalization-group techniques.\cite{Belitz_Kirkpatrick_1997, Belitz_Kirkpatrick_2014}  In 
contrast to the clean case, the prefactor ${\tilde a}_d$ is nonzero already at first order in the 
spin-triplet interaction amplitude.

\subsubsection{Soft modes and the single-particle spectrum}
\label{subsubsec:II.A.3}

The preceding arguments imply that one can deduce the soft-mode spectrum from the field
dependence of the of the spectrum of the single-particle Hamiltonian. Suppose the energy
eigenvalues are $E_{\bm k}^{\beta}$, with $\beta$ an index that describes spin and/or
other degrees of freedom. Then the quasiparticle resonances are
\bse
\label{eqs:2.12}
\be
F_k^{\beta} = \frac{1}{i\omega_m - E_{\bm k}^{\beta}}
\label{eq:2.12a}
\ee
and the sheets of the Fermi surface are defined as the set of ${\bm k}$ for which
\be
E_{\bm k}^{\beta} = \mu\ .
\label{eq:2.12b}
\ee
\ese
In the example in Sec.~\ref{subsubsec:II.A.1}, $\beta = \sigma$
and $E_{\bm k}^{\beta} = \epsilon_{\bm k} - \beta h$. The existence of a SM1 of the form
shown in Eqs.~(\ref{eqs:2.5}) is equivalent to the statement that two energy sheets
$E_{\bm k}^{\beta} = \mu$ and $E_{\bm k}^{\beta'} = \mu$ wth $\beta' \neq \beta$ or, equivalently,
the corresponding Fermi surfaces, are
degenerate in zero field but not in a nonzero field. This property of the noninteracting
Hamiltonian suffices for predicting certain properties of the interacting system, including
the nature of the quantum phase transition whose order parameter is conjugate to the
field, see Sec.~\ref{subsec:II.B} below. 

\begin{figure}[t]
\includegraphics[width=8.5cm]{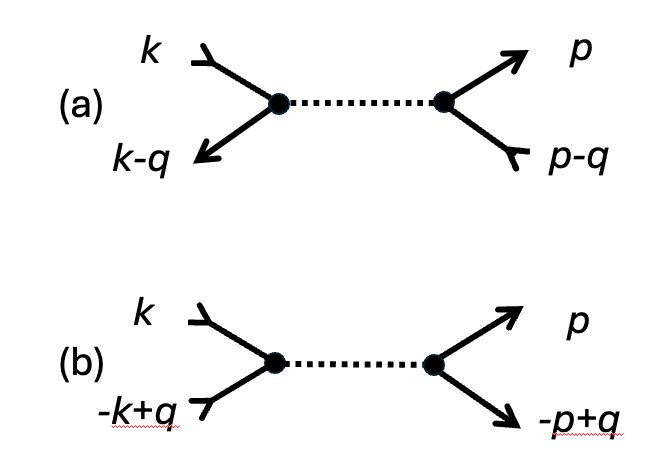}
\caption{Electron-electron interaction in (a) the particle-hole channel and (b) the particle-particle channel.}
\label{fig:0}
\end{figure}

For the mode defined in Eq.~(\ref{eq:2.5a}), the interaction that provides the 
frequency mixing that is crucial for coupling the soft mode to the relevant observable is
the interaction in the particle-hole channel, which is shown diagrammatically in Fig.~\ref{fig:0}(a).
For later reference we also introduce modes in the particle-particle channel defined by
\be
\psi_{i\omega_m}^{\beta\beta'}({\bm q},i\Omega_n) = \frac{1}{V} \sum_{\bm k} F_k^{\beta} F_{-k+q}^{\beta'}\ ,
\label{eq:2.13}
\ee
which couple to observables via the interaction in the particle-particle channel shown in
Fig.~\ref{fig:0}(b).
For these modes to be soft of the first kind the relevant criterion is that $E_{\bm k}^{\beta}$ and
$E_{-{\bm k}}^{\beta'}$ are degenerate in zero field with a nonzero field lifting the degeneracy. 
For systems that are invariant under spatial inversion this is the same criterion as in the
particle-hole channel. As we will discuss in Sec.~\ref{subsubsec:III.A.4}, it further turns out that the 
effects of the soft modes in the particle-particle channel are logarithmically weaker than those
of the soft modes in the particle-hole channel. Consequently, the modes defined by Eq.~(\ref{eq:2.13})
become relevant only if spatial inversion invariance is broken {\em and} there are no soft
modes in the particle-hole channel. This is realized in non-centrosymmetric ferromagnets with a strong 
spin-orbit interaction, see Sec.~\ref{subsubsec:III.A.4}.

\subsection{Consequences for phase transitions}
\label{subsec:II.B}       

Now consider a phase transition for which $\mathfrak{O}$ is the order parameter, so $\langle\mathfrak{O}\rangle = 0$
in the disordered phase, and $\langle\mathfrak{O}\rangle \equiv \Phi \neq 0$ in the ordered phase. (If ${\mathfrak O}$
is modulated, then $\Phi = \langle {\mathfrak A}\rangle$, with ${\mathfrak A}$ the amplitude of ${\mathfrak O}$.) Then 
$\Phi$ will enter the action the same way as the conjugate field $\mathfrak{h}$. The standard Landau theory of phase
transitions ignores order-parameter fluctuations and formally integrates out all degrees of freedom other then $\Phi$,
so the partition function is
\bse
\label{eqs:2.14}
\be
Z[\Phi] = \int D[\bar\psi,\psi]\,e^{S_0 + S_{\text{int}}}\,e^{S_{\text{c}}[\Phi;\bar\psi,\psi]}
\label{eq:2.14a}
\ee
where
\be
S_{\text{c}}[\Phi;\bar\psi,\psi] = c\,\Phi \int dx\,\mathfrak{O}(x)
\label{eq:2.14b}
\ee
Here $c$ is a coupling constant, and $x\equiv({\bm x},\tau)$ comprises both position and imaginary time,
with $\int dx = \int_V d{\bm x} \int_0^{1/T} d\tau$. The free energy density now is
\be
f(\Phi) = -(T/V) \ln Z[\Phi]
\label{eq:2.14c}
\ee
\ese

The standard Landau theory assumes that $f$ is an analytic function of $\Phi$. The discussion in 
Sec.~\ref{subsubsec:II.A.2} implies that the latter assumption fails in the presence of any modes that 
are soft of the first kind with respect to $\mathfrak{h}$, and hence to $\Phi$. In order to see this
explicitly, consider the $\mathfrak{O}$-susceptibility in the presence of the conjugate field
$\mathfrak{h}$, which is given by
\be
\chi_{\mathfrak{O}}(\mathfrak{h}) = \frac{T}{V} \int dx\,dy\ \langle \delta\mathfrak{O}(x)\,\delta\mathfrak{O}(y)\rangle_{S_0 + S_{\text{int}} + S_{\mathfrak{h}}}\ .
\label{eq:2.15}
\ee
Here $\delta\mathfrak{O}(x) = \mathfrak{O}(x) - \Phi$, and $\langle \ldots \rangle_S$ denotes an average
with respect to an action $S$. Differentiating Eq.~(\ref{eq:2.14c}) twice with respect to $\Phi$ yields
\be
d^2 f/d\Phi^2 = (V/T) c^2 \chi_{\mathfrak{O}}(\mathfrak{h}=c\Phi)\ ,
\label{eq:2.16}
\ee
or
\be
f(\Phi) = \frac{-V}{T}\,c^2 \int_0^{\Phi} d\Phi_1 \int_0^{\Phi_1} d\Phi_2\, \chi_{\mathfrak{O}}(\mathfrak{h}=c\,\Phi_2)\ .
\label{eq:2.17}
\ee
Now the analytic contributions to $\chi_{\mathfrak{O}}$ will produce a standard Landau free energy that is
analytic in $\Phi$, but the nonanalytic contribution discussed in Sec.~\ref{subsubsec:II.A.2} will produce a 
nonanalytic contribution to $f$. The generalized Landau free energy function that takes into account the
fermionic soft modes of the first kind then has the form
\be
f_{\text{gen}}(\Phi) = -\mathfrak{h} \Phi + \frac{r}{2}\,\Phi^2 + \frac{u}{4}\,\Phi^4 + O(\Phi^6) + \delta f(\Phi)
\label{eq:2.18}
\ee
where $r$ and $u$ are the parameters of the ordinary Landau theory, and $\delta f$ is given by 
Eq.~(\ref{eq:2.17}) with the integrand given by the nonanalytic contribution to $\chi_{\mathfrak{O}}$. 
If the nonanalyticity is strong enough to dominate the $\Phi^4$ term in the ordinary Landau theory 
(this depends on how the field-induced 
mass of the SM1 scales with the field), the presence of $\delta f$ can lead either a first-order transition 
(for $u>0$, i.e., in situations where ordinary Landau theory predicts a second-order transition), or a 
second-order transition with exponents that are different from the mean-field exponents of ordinary 
Landau theory. The former case is a quantum analog of the classical fluctuation-induced first-order 
transitions in superconductors and liquid crystals.\cite{Halperin_Lubensky_Ma_1974}

The generalized Landau free energy given by Eq.~(\ref{eq:2.18}) ignores order-parameter fluctuations,
and we need to discuss the validity of this approximation. For classical second-order transitions, Landau 
theory in general does not give the correct critical behavior in spatial dimensions $d$ less or equal than
the upper critical dimension $d_{\text{c}} = 4$. This is because order-parameter fluctuations become more
and more important as the critical point is approached and modify the critical behavior.\cite{Wilson_Kogut_1974, Ma_1976}
A classical fluctuation-induced first-order transition often survives the order-parameter fluctuations, but the latter
are still strong and make the first-order transition weak and hard to observe.\cite{Anisimov_et_al_1990}
In a quantum system the effective dimension is given by $d_{\text{eff}} = d + z$, with $z$ an appropriate
dynamical exponent.\cite{Hertz_1976} As a result, the system in the absence of the coupling of the order
parameter to the fermionic soft modes is usually above its critical dimension. This means that a first-order
transition induced by $\delta f$ is much more robust than a classical fluctuation-induced first-order 
transition, and in the case of a second-order transition the exponents predicted by the generalized Landau
theory, Eq.~(\ref{eq:2.18}), tend to be exact. One caveat is the fact that there often are multiple dynamical
exponents, and their interplay can lead to coupling constants that are marginal (in the renormalization-group
sense) even though they are irrelevant by power counting. An example is the quantum ferromagnetic transition
in disordered metals, where such a marginal operator puts the 3-d system at its upper critical dimension,
even though all dynamical exponents are greater than 1. This leads to complicated logarithmic corrections
to the scaling behavior predicted by the generalized Landau theory, see Sec.~\ref{subsubsec:III.A.1}.

We will discuss various magnetic order parameters and their conjugate fields. In all cases the crucial
question with respect to the nature of the quantum phase transition is the existence of soft modes of
the first kind.

\section{Implications for magnetic transitions}
\label{sec:III}

In this section we explain how the concepts described in the previous section apply to various
quantum magnetic transitions. For all of the transitions that we will consider, the nonmagnetic or
paramagnetic phase is a Fermi liquid that may be of Landau or Dirac type, which differ with respect
to both the single-particle excitations and the two-particle correlation functions and the related soft
modes. In some cases the Fermi liquid is modified by a spin-orbit interaction. The soft-mode
characteristics of the various types of magnets are summarized in Table~\ref{table:I}.

\subsection{Ferromagnetic systems}
\label{subsec:III.A}

In ferromagnets, the relevant observable ${\frak O}$ is the electronic spin density, the order parameter $\Phi$
is the homogeneous magnetization $m$, and the conjugate field ${\frak h}$ is the physical magnetic field ${\bm h}$.
For the purposes of our classification, this class also includes systems where a more complicated spin texture
results in a homogeneous magnetization, as is the case in ferrimagnets and canted antiferromagnets. 

\begin{table*}[t]
\centering
\caption{Types of metallic magnets, classified by the nature of the nonmagnetic state, the presence of a
              homogeneous magnetization, a field-split Fermi surface, soft modes, and order of the quantum
              phase transition in clean 3-d systems.
              FL = Fermi liquid, LFL = Landau Fermi liquid, DFL = Dirac Fermi liquid, FS = Fermi surface, 
              SM1 = Soft mode of the first kind (see text), LR = long range, QPT = quantum phase transition,
              FM = ferromagnet, AFM = antiferromagnet, HM = helimagnet}
\vskip 2pt
\begin{ruledtabular}
\begin{tabular}{l c c  c c c c}
Type of magnet
   & Type of
      & homogeneous
              & field-split
                  & SM1 \& LR
                       & ordering
                           & Order 
\\

   & FL
      & magnetization
                   & FS
                       & correlations
                           & wave number 
                                & of QPT
\\
\\ [-7pt]
\hline\\[-8pt]
\hline\\[-3pt]
Landau FM$^{\,(a,b)}$
   & LFL
      & yes
                  & yes
                       & yes
                            & 0
                                & 1$^{\text{st}}$
\\
\\[-7pt]
Ferrimagnet
    & LFL
       & yes
                  & yes
                       & yes 
                            & 0
                                & 1$^{\text{st}}$                       
\\
 \\[-7pt]
Canted AFM
    & LFL
       & yes
                & yes
                     & yes 
                          & 0
                               & 1$^{\text{st}}$                     
\\
 \\[-7pt]
\hline\\ [-3pt]
Dirac FM$^{(c)}$
   &  DFL
      & yes
                    & yes
                        & \ \ \ yes$^{(d)}$
                             & 0
                                  & 1$^{\text{st}}$                        
\\
\\[-7pt]
\hline\\ [-3pt]
Magnetic Nematic
   &  LFL
      & no
                    & yes
                        & yes
                             & 0
                                  & 1$^{\text{st}}$                        
\\
 \\[-7pt]
\hline\\ [-3pt]
Altermagnet
   &  LFL
      & no
                    & yes
                        & yes
                             & \ \ \ 0$^{\,\text{(e)}}$
                                  & 1$^{\text{st}}$                        

\\
 \\[-7pt]
\hlineB{3.0}
 \\[-4pt]
Helimagnet
   &  \ \ \ LFL$^{(f)}$
      & no
                    & yes
                        & yes 
                             & $\neq 0$
                                  & \ \ \ 2$^{\text{nd}\, \text{(g)}}$                       
\\
 \\[-7pt]
Magnetic Smectic
   & LFL
         & no
                    & yes
                        & yes 
                             & $\neq 0$
                                  & \ \ \ 2$^{\text{nd}\, \text{(g)}}$                       
\\
 \\[-7pt]
N{\'e}el AFM
   &  LFL
      & no
                    & yes
                        & yes 
                             & $\neq 0$
                                  & \ \ \ 2$^{\text{nd}\,\text{(g)}}$                       
\\
 \\[-7pt]
\hlineB{3.0}
\\ [-4pt]
non-centrosymmetric
   & 
      & 
                  & 
                       & 
                           & 
\\
\quad FM with strong
   &  LFL$^{(f)}$
      & yes
                    & \ \ no$^{\text{(h)}}$
                        & \ \ \ yes$^{\text{(i)}}$
                             & 0
                                  & \ \ \ 2$^{\text{nd}\,\text{(j)}}$                        
\\
\quad spin-orbit coupling
   &  
      & 
                    & 
                        & 
                           &                  
\\
 \\[-7pt]
 \\[-9pt]
\hline\hline\\[-5pt]
\multicolumn{6}{l} {$^{(a)}$ Based on LFL, negligible spin-orbit coupling.\quad 
                             }\\
\multicolumn{6}{l} {$^{(b)}$ Any spin symmetry (Ising, XY, Heisenberg); includes canted FMs.}\\
\multicolumn{6}{l} {$^{(c)}$  Based on DFL. 
                             }           \\      
\multicolumn{6}{l} {$^{(d)}$  Different from soft modes in generic FMs.}\\   
\multicolumn{6}{l} {$^{(e)}$  For the nematic part of the OP. 
                              }\\     
\multicolumn{6}{l} {$^{(f)}$  LFL modified by spin-orbit coupling.}\\                                   
\multicolumn{6}{l} {$^{(g)}$ Field nonanalyticity induced by SM1 not strong enough to change the order of the QPT.}\\
\multicolumn{6}{l} {$^{(h)}$  Energy surfaces $E_{\bm k}^{\sigma}$ and $E_{-{\bm k}}^{\sigma}$ are field-split and lead to a SM1.
                              }\\
\multicolumn{6}{l} {$^{(i)}$  SM1 in the particle-particle channel only.}\\
\multicolumn{6}{l} {$^{(j)}$  In 3-d systems.}\\   
\end{tabular}
\end{ruledtabular}
\vskip -0mm
\label{table:I}
\end{table*}

\subsubsection{Ferromagnets based on Landau Fermi liquids}
\label{subsubsec:III.A.1}

\begin{figure}[h]
\includegraphics[width=8.5cm]{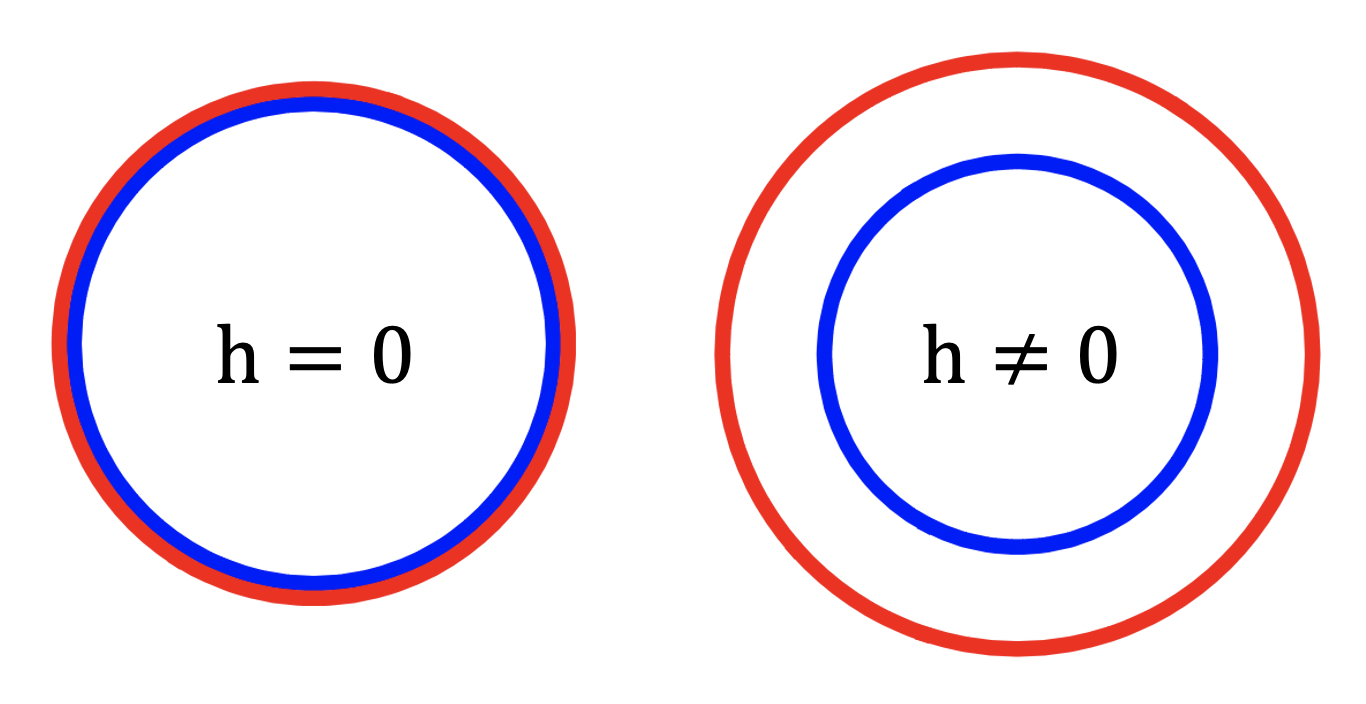}
\caption{Field splitting of the Fermi surface in a FM based on a Landau Fermi liquid. The sheets
of the Fermi surface are labeled by the spin projection. }
\label{fig:1}
\end{figure}

The simplest case is a ferromagnet based on a simple, or Landau, metal whose conduction electrons form an
ordinary Landau Fermi liquid. This is the simple example we already discussed in Sec.~\ref{subsubsec:II.A.1};
the single-particle Hamiltonian is given by Eq.~(\ref{eq:2.1a}). The quasiparticle resonances are determined by the
eigenvalues of the Hamiltonian; they are given by Eq.~(\ref{eq:2.3}). The single-particle spectrum in zero field 
is two-fold Kramers degenerate. This degeneracy is lifted by both a nonzero field and a nonzero magnetization, 
see Eq.~(\ref{eq:2.3}) and Fig.~\ref{fig:1}. 
The resulting spin-split Fermi surface leads to the soft modes shown in Eq.~(\ref{eq:2.5c}), which are of the 
first kind with respect to $h$ or $m$ if the two spin projections are different, $\sigma \neq \sigma'$. In clean 
systems, the nonanalytic part of the order-parameter susceptibility is given by Eq.~(\ref{eq:2.9}), and 
Eq.~(\ref{eq:2.17}) leads to
\be
\delta f(m) = -v_d \times \begin{cases} m^{d+1} & \text{for}\ 1<d<3 \\
                                                             m^4 \ln(1/m^2) & \text{for}\ d=3
                                      \end{cases}         
\label{eq:3.1}
\ee
with $v_d>0$. The complete generalized Landau free energy now is, for $d=3$,
\be
f_{\text{gen}}(m) = -h\,m + \frac{r}{2}\,m^2 - \frac{v}{4}\, m^4 \ln(1/m^2) + \frac{(u+v/2)}{4}\, m^4\ .
\label{eq:3.2}
\ee
Since $v = 4v_3 > 0$, this leads to a first-order transition, as was first discussed in 
Ref.~\onlinecite{Belitz_Kirkpatrick_Vojta_1999}.

\begin{figure}[t]
\includegraphics[width=7.5cm]{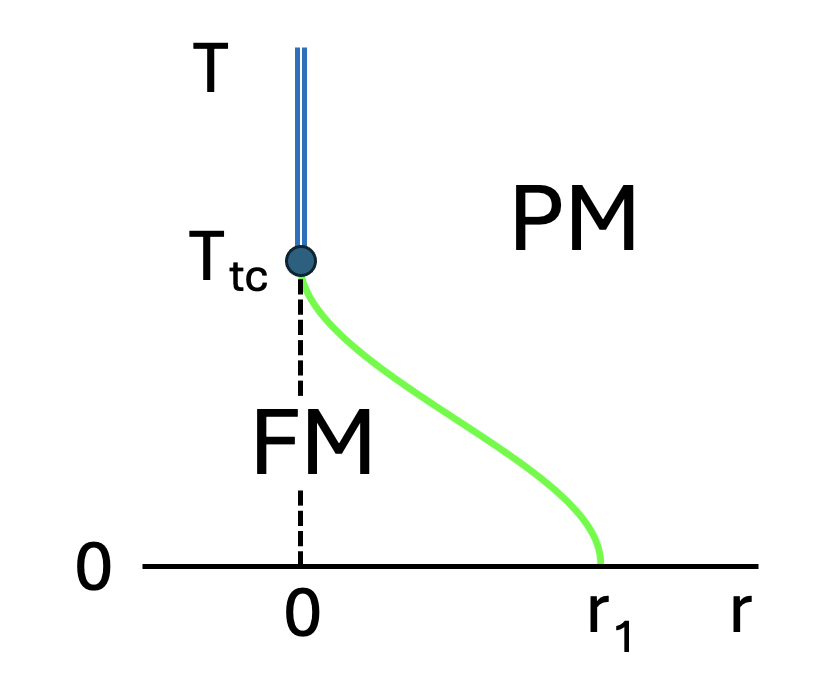}
\caption{Schematic phase diagram in the $T$-$r$ plane showing the boundary between the ferromagnetic (FM)
              and the paramagnetic (PM) phase. A line of second-order transitions (blue double line)
              meets a line of first-order transition (green single line) at the tricritical point $(T_{\text{tc}},r=0)$.
              The quantum phase transition is of first order and located at $(T=0, r_1)$.}
\label{fig:2}
\end{figure}

A nonzero temperature $T>0$ gives the soft modes a mass, which cuts off the logarithmic singularity
in the free energy. Ignoring a term that is of no qualitative importance for the phase diagram, the 
generalized Landau equation of state becomes
\bse
\label{eqs:3.3}
\be
h = r\,m - v\,m^3 \ln\left(\frac{1}{m^2 + T^2}\right) + u\,m^3
\label{eq:3.3a}
\ee
with $m$, $h$, and $T$ measured in natural units. With increasing temperature the
first-order transition weakens, and there is a tricritical point at $T_{\text{tc}} = e^{-u/2v}$. The
schematic phase diagram for $h=0$ is shown in Fig.~\ref{fig:2}.  
In a nonzero magnetic field the phase diagram shows tricritical wings.\cite{Belitz_Kirkpatrick_Rollbuehler_2005}
In 2-d the nonanalyticity is stronger and the equation of state becomes, at zero temperature,
\be
h = r\,m - v\,m^2 + u\,m^3\ 
\label{eq:3.3b}
\ee
\ese

Quenched disorder also gives the soft modes in Eq.~(\ref{eq:2.5c}) a mass, as does a nonzero temperature,
but it also leads to the emergence of new soft modes. For weak disorder, when the latter can be ignored,
the 3-d equation of state becomes\cite{Sang_Belitz_Kirkpatrick_2014}
\be
h = r\,m - v\,m^3 \ln\left(\frac{1}{m^2 + (T + \rho)^2}\right) + u\,m^3\ ,
\label{eq:3.4}
\ee
with $\rho$ a dimensionless measure of the disorder. With increasing disorder the tricritical temperature
thus decreases until it vanishes at a critical value of the disorder, see Fig.~\ref{fig:3} in Sec.~\ref{subsubsec:III.A.4} 
below. This has been observed in 
various materials.\cite{Huang_et_al_2016, Goko_et_al_2017, Mishra_et_al_2020}
In the opposite limit of strong 
disorder the physics is dominated by the emergent soft modes related to the diffusive electron dynamics, 
and the nonanalytic contribution to the order-parameter susceptibility
is given by Eq.~(\ref{eq:2.11}). Via Eqs.~(\ref{eq:2.17}, \ref{eq:2.18}) this leads, for $d=3$, to 
\be
f_{\text{gen}}(m) = -h\,m + r\,m^2 + v\,m^{5/2} + u\,m^4
\label{eq:3.5}
\ee
with $v>0$. This describes a second-order transition with non-mean-field exponents.\cite{Kirkpatrick_Belitz_1996}
The crossover between the weak and strong disorder regimes has been discussed in
Ref.~\onlinecite{Sang_Belitz_Kirkpatrick_2014}. 2-d systems in the disordered case are complicated
due to electron localization effects.

As we mentioned at the end of Sec.~\ref{subsec:II.B}, this case provides an example where 
order-parameter fluctuations modify the critical behavior predicted by the generalized Landau
theory, if only via logarithmic corrections to scaling. At first sight this seems surprising: The two
dynamical critical exponents in the theory are a diffusive dynamical exponent $z_{\text{diff}} = 2$
that describes the dynamics of the diffusive electrons, and a critical dynamical exponents 
$z_{\text{c}} = d$ (for $2<d<4$) that describes the dynamics of the order-parameter fluctuations.
It therefore seems that the effective dimensionality $d_{\text{eff}}= d+z$ is safely above the upper
critical dimension $d_{\text{uc}} = 4$ irrespective of which dynamical exponent enters the effective
dimensionality. However, since the renormalization-group scale factor can represent either a
wave number or a frequency, naive power counting cannot, in certain contexts, distinguish
between a scale dimension equal to $d-2$ and one equal to zero.\cite{Belitz_et_al_2001a, Belitz_et_al_2001b}
The net effect is equivalent to the existence of a dynamical exponent $z = 2 = 4 - d + (d-2)$ that
turns into an effective dynamical exponent $z_{\text{eff}} = 4 - d$. This in turn makes the effective
dimensionality equal to $d_{\text{eff}} = d + z_{\text{eff}} = 4$, which equals the upper critical
dimension for all physical spatial dimensions $2<d<4$, which leads to logarithmic corrections to
scaling. For details regarding this mechanism, see Refs.~\onlinecite{Belitz_et_al_2001a, Belitz_et_al_2001b}.
The logarithmic corrections to scaling are hard to observe and mimic power laws with effective
exponents in large regions of parameter space.\cite{Kirkpatrick_Belitz_2014}

Several comments are in order here. In the above example, the homogeneous magnetization
leads to a spin-split Fermi surface in the ordered phase, but that is not the salient point. As
explained in Sec.~\ref{subsubsec:II.A.2}, it is the existence of a soft mode of the first kind 
that leads to the free-energy functional being a nonanalytic function of the order parameter,
i.e., a mode that is soft in the absence of both a nonzero order parameter and the conjugate
field, but becomes massive in the presence of the field. The structure of the soft mode, 
Eqs.~(\ref{eqs:2.5}), implies that the existence of degenerate sheets of the Fermi surface 
that are split by a nonzero field suffices for producing the effect. That is, the underlying 
long-ranged correlations are a property of the non-magnetic state, in this example, the 
Landau Fermi liquid where spin correlations are long-ranged as described by 
Eqs.~(\ref{eqs:1.2}). We also stress again that the soft modes are the result of very general
symmetries that are present in both noninteracting and interacting electron systems. Therefore, 
considering noninteracting electrons suffices for determining the soft-mode structure. However,
the presence of electron-electron interactions is crucial for the frequency mixing that produces 
the nonanalyticities from the soft modes: in noninteracting systems the soft modes are present,
but the nonanalyticities are not. We also stress that these arguments are independent of the
origin of the magnetism, and are valid irrespective of whether the magnetic moments are supplied by 
the conduction electrons (itinerant ferromagnets) or by electrons in a different band. They are
further independent of the symmetry in spin space, and hence are equally valid for Heisenberg,
XY, and Ising ferromagnets, and also for canted ferromagnets, where the moments are not
collinear.

\subsubsection{Ferrimagnets, and canted antiferromagnets}
\label{subsubsec:III.A.2}

The arguments given in Sec.~\ref{subsubsec:III.A.1} also hold if the homogeneous magnetization 
is the result of a more complicated spin texture. An example is provided by ferrimagnets, i.e.,
antiferromagnets where the moments on the two sublattices do not completely compensate one 
another, which results in a net magnetization. Another one is given by canted antiferromagnets, 
where the same effect is achieved by the moments on the sublattices not being collinear. 

For all of these systems the magnetic phase transition is described by the generalized
Landau free energy given in Eqs.~(\ref{eq:3.2}) or (\ref{eq:3.5}), provided the single-particle
Hamiltonian is given by Eq.~(\ref{eq:2.1a}).\cite{Kirkpatrick_Belitz_2012}

\subsubsection{Ferromagnets based on Dirac Fermi liquids}
\label{subsubsec:III.A.3}

Different kinds of soft modes exist in metals where a strong spin-orbit interaction of the form
${\bm\sigma}\cdot{\bm k}$ leads to a linear crossing of two bands.\cite{Abrikosov_Beneslavskii_1970}
Such a term can appear with either sign, which leads to electron species with two different chiralities.
In systems that are invariant under both time reversal (in the absence of a magnetic field or a 
magnetization) and spatial inversion the two chiralities mix in a symmetric way, and the most
general single-particle Hamiltonian has the form\cite{Zhang_et_al_2009, Liu_et_al_2010}
\begin{align}
{\mathcal H}_{\bm k} &= \xi_{\bm k}\,(\pi_0\otimes\sigma_0) + \vso(\pi_3\otimes{\bm\sigma})\cdot{\bm k} +\Delta(\pi_1\otimes\sigma_0) 
\nonumber\\
& \hskip 10pt - h(\pi_0\otimes\sigma_3) \ .
\label{eq:3.6}
\end{align}
Here $(\pi_1,\pi_2,\pi_3)$ is a second set of Pauli matrices that describes the chirality degree
of freedom, and $\pi_0 = \sigma_0$. $\vso$ is the coupling constant for the spin-orbit interaction,
$\Delta$ is the coupling constant for the chirality-mixing term, and $h$ is a magnetic field in
3-direction. Due to the cone structure that results from the linear crossing of the bands
such systems are called Dirac metals, and we refer to the underlying Fermi liquid as a
Dirac Fermi liquid. 
\begin{figure}[b]
\includegraphics[width=8.5cm]{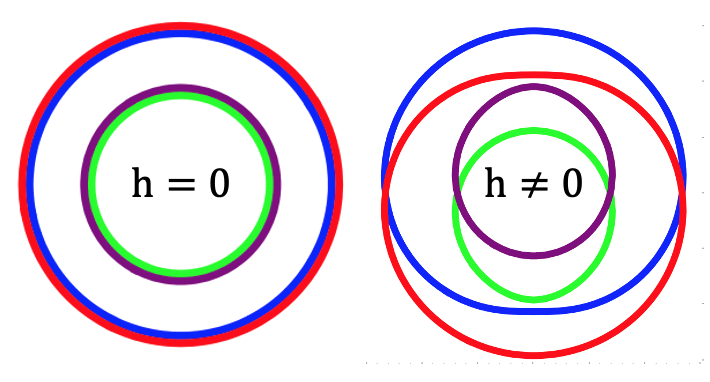}
\caption{Field splitting of the Fermi surface in a FM based on a Dirac Fermi liquid. The sheets
of the Fermi surface are labeled by the chirality index $\alpha$ and the Stoner splitting index
$\sigma$, not by the spin projection. }
\label{fig:3}
\end{figure}

Systems described by such a Hamiltonian have received much attention because of their
topological properties,\cite{Wan_et_al_2011, Yang_Lu_Ran_2011, Burkov_Balents_2011}
but this is not relevant for the soft-mode structure. ${\mathcal H}_{\bm k}$ is now a $4\times 4$ 
matrix, so the single-particle spectrum has four branches and in general is quite complicated. Its
soft-mode properties have been analyzed in Ref.~\onlinecite{Kirkpatrick_Belitz_2019a}. 
Putting $\Delta = 0$, which has no bearing on the qualitative aspects of the soft-mode spectrum, 
the quasiparticle resonances that generalize Eq.~(\ref{eq:2.3}) become
\be
F_k^{\alpha\sigma} = \frac{1}{i\omega_m - \xi_{\bm k} - \sigma\vert\alpha \vso{\bm k} - {\bm h}\vert}
\label{eq:3.7}
\ee
Here $\alpha = \pm$ is the chirality index; together with $\sigma = \pm$ it labels the four branches
of the spectrum. Since spin is not a good quantum number due to the spin-orbit interaction, 
$\sigma$ can be interpreted as representing two different spin projections only in the limit $\vso=0$.
The Green function, within these approximations, is diagonal in the chirality index and the
diagonal elements (which are still $2\times 2$ matrices), have the form
\be
{\mathcal G}_k^{\alpha} = \sum_{\sigma=\pm} F_k^{\alpha\sigma} {\mathcal M}^{\sigma}(\sigma\alpha\hat{\bm k})
\label{eq:3.8}
\ee
with ${\mathcal M}^{\sigma}$ from Eq.~(\ref{eq:2.4b}). In zero field the four branches of
the Fermi surface are pairwise degenerate. A nonzero field splits these degeneracies,
see Fig.~\ref{fig:3}. 

The structure of Eqs~(\ref{eq:2.5a}, \ref{eq:2.5b})
is unchanged, and Eq.~(\ref{eq:2.5c}) gets generalized to
\begin{widetext}
\bse
\label{eqs:3.9}
\be
 \phi_{\alpha \alpha'}^{\sigma \sigma'}({\bm q}, i\Omega_n) = \NF
   \int \frac{d\Omega_{\bm k}}{4\pi}\,\frac{}{i\Omega_n - \vF\hat{\bm k}\cdot{\bm q} + \sigma' \vert \alpha' \vso(\kF\hat{\bm k}+{\bm q}) - {\bm h}\vert - \sigma\vert \alpha \vso\kF\hat{\bm k} - {\bm h}\vert}
\label{eq:3.9a}
\ee
Now consider the modes with $\sigma' = -\sigma$, i.e., the chiral generalizations of the relevant soft modes in the
Landau Fermi liquid case. We see that the spin-orbit interaction gives these modes a mass, irrespective of the
values of $\alpha$ and $\alpha'$: For $h\ll \vso\kF$ we have
\be
\phi_{\alpha \alpha'}^{\sigma, -\sigma}({\bm q}, i\Omega_n; i\omega_m) = \NF
   \int \frac{d\Omega_{\bm k}}{4\pi}\,\frac{1}
   {i\Omega_n - \vF\hat{\bm k}\cdot{\bm q} - 2\sigma \vso\kF}
\label{eq:3.9c}
\ee
However, the modes with $\sigma = \sigma'$ remain soft,
\be
\phi_{\alpha \alpha'}^{\sigma\sigma}({\bm q}, i\Omega_n; i\omega_m) = \NF
   \int \frac{d\Omega_{\bm k}}{4\pi}\,\frac{1}
   {i\Omega_n - \vF\hat{\bm k}\cdot{\bm q} + \sigma(\alpha - \alpha')\hat{\bm k}\cdot{\bm h}}
\label{eq:3.9d}
\ee
\ese
\end{widetext}
For $\alpha \neq \alpha'$ these modes are of the first kind with respect to $h$, in agreement with the field-splitting
of the Fermi surface, see Fig.~\ref{fig:3}.
The spin-orbit interaction thus invalidates the mechanism that generates a soft mode of the first kind in a Landau
Fermi liquid, but it also leads to emergent chiral degrees of freedom which create a new mechanism.  Accordingly,
the ferromagnetic transition in a clean Dirac metal is generically of first order, and the generalized Landau free
energy density is the same as in a Landau metal and given by Eq.~(\ref{eq:3.2}).\cite{Belitz_Kirkpatrick_2019}
What underlies this result is the fact that spin correlations in a Dirac Fermi liquid are long-ranged and given by
Eqs.~(\ref{eqs:1.2}), as is the case in a Landau Fermi liquid, despite the fact that the relevant soft modes are
of a very different nature.\cite{Kirkpatrick_Belitz_2019a}

\subsubsection{Non-centrosymmetric ferromagnets with strong spin-orbit coupling}
\label{subsubsec:III.A.4}

In all of the examples discussed so far the single-particle Hamiltonian has been invariant under spatial
inversions. In the Landau ferromagnet case the single-particle energy eigenvalues obey
\bse
\label{eqs:3.10}
\be
E_{-{\bm k}}^{\sigma} = E_{\bm k}^{\sigma}\ ,
\label{eq:3.10a}
\ee
and in the Dirac ferromagnet case,
\be
E_{-{\bm k}}^{\sigma,\alpha} = E_{\bm k}^{\sigma,-\alpha}\ .
\label{eq:3.10b}
\ee
\ese
It therefore sufficed to consider the soft modes in the particle-hole channel, Eq.~(\ref{eq:2.5a}).
A qualitatively different situation arises in systems that have the same spin-orbit coupling as in
Sec.~\ref{subsubsec:III.A.3}, but lack spatial inversion symmetry, so the particle-particle channel,
Eq.~(\ref{eq:2.13}), needs to be considered. In this
case no emergent degrees of freedom arise, and the single-particle Hamiltonian is given by, instead of
Eq.~(\ref{eq:3.6}),\cite{Dyakonov_2008, so_footnote}
\be
{\mathcal H}_{\bm k} = \xi_{\bm k}\,\sigma_0 + \vso\,{\bm\sigma}\cdot{\bm k}  - {\bm h}\cdot{\bm\sigma}\ .
\label{eq:3.11}
\ee
The quasiparticle resonances now are
\be
F_k^{\sigma} = \frac{1}{i\omega_m - \xi_{\bm k} - \sigma \vert \vso{\bm k} - {\bm h}\vert}\ ,
\label{eq:3.12}
\ee
and the Green function is
\be
{\mathcal G}_k = \sum_{\sigma=\pm} F_k^{\sigma} {\mathcal M}^{\sigma}
                                                  \left(\frac{\vso{\bm k}-{\bm h}}{\vert \vso{\bm k}-{\bm h}\vert}\right)\ .
\label{eq:3.13}
\ee

\paragraph{Soft modes in the particle-hole channel}
\label{par:III.A.4.a}

\begin{figure}[t]
\includegraphics[width=8.5cm]{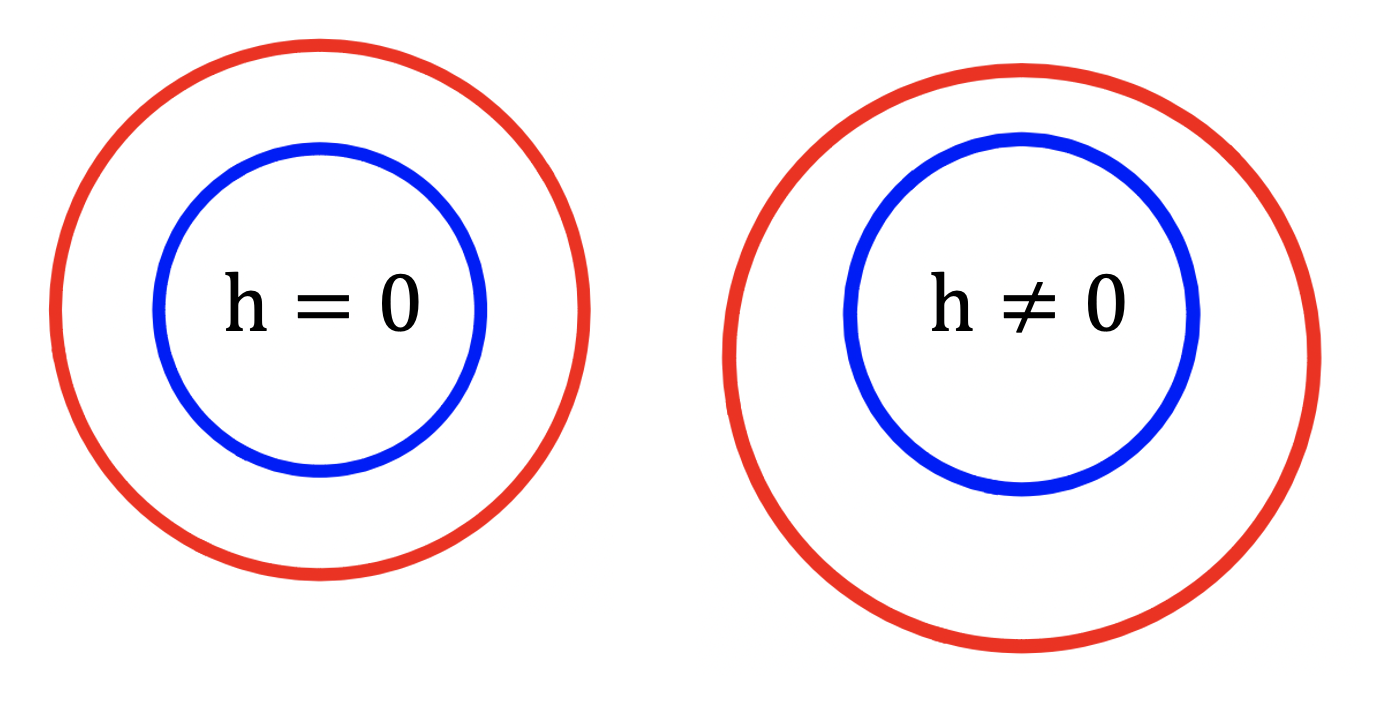}
\caption{Schematic Fermi surface of a non-centrosymmetric FM with a strong spin-orbit interaction. The sheets
cannot be labeled by the spin projection. }
\label{fig:4}
\end{figure}

The spin-orbit interaction splits the Fermi surface even in zero field and there are no degeneracies that
a nonzero field could lift, see Fig.~\ref{fig:4}.
The analog of Eq.~(\ref{eq:2.5c}) becomes
\bea
\phi^{\sigma\sigma'}({\bm q},i\Omega_n) &=&
\nonumber\\
&& \hskip -50pt \NF \int \frac{d\Omega_{\bm k}}{4\pi}\,\frac{1}{i\Omega_n - \vF\hat{\bm k}\cdot{\bm q} 
                       - (\sigma - \sigma') \vert \vso\kF\hat{\bm k} - {\bm h}\vert}\ .
   \nonumber\\
\label{eq:3.14}
\eea

We see that the $\phi^{\sigma\sigma'}({\bm q},i\Omega_n)$ with $\sigma = \sigma'$ are soft, but a magnetic field does
not give them a mass, so they are soft of the second kind with respect to $h$. The modes with $\sigma \neq \sigma'$,
on the other hand, are massive even for $h=0$. The conclusion is that in the particle-hole channel there are no soft 
modes of the first kind in hese systems, consistent with the structure of the Fermi surface. 

\medskip
\paragraph{Soft modes in the particle-particle channel}
\label{par:III.A.4.b}

The energy eigenvalues
\bse
\label{eqs:3.15}
\be
E_{\bm k}^{\sigma}({\bm h}) = \epsilon_{\bm k} - \sigma \vert \vso{\bm k} - {\bm h}\vert
\label{eq:3.15a}
\ee
have the property 
\be
E_{-{\bm k}}^{\sigma}({\bm h}) = E_{{\bm k}}^{\sigma}(-{\bm h}) 
\label{eq:3.15b}
\ee
\ese
Accordingly, $E_{\bm k}^{\sigma}$ and $E_{-{\bm k}}^{\sigma}$ are degenerate for ${\bm h} = 0$,
but not for ${\bm h} \neq 0$, and since there is no SM1 in the particle-hole channel we need to
consider the particle-particle channel. The mode defined by Eq.~(\ref{eq:2.13}) is, for $h\ll \vso\kF$,
\bea
\psi_{i\omega_m}^{\sigma\sigma'}({\bm q},i\Omega_n) &=& 
\nonumber\\
&&\hskip -65pt \frac{1}{2i\omega_m - i\Omega_n - \vF\hat{\bm k}\cdot{\bm q} 
     - (\sigma - \sigma')\vso\kF + (\sigma + \sigma')\hat{\bm{k}}\cdot{\bm h}}\ \ 
     \nonumber\\
\label{eq:3.16}
\eea
We see that for $\sigma = \sigma'$ this mode is a SM1.\cite{Zak_Maslov_Loss_2010, Miserev_Loss_Klinovaja_2022,
Belitz_Kirkpatrick_2026, Miserev_Loss_Klinovaja_2026, disorder_footnote}

\medskip
\paragraph{Nature of the quantum phase transition}
\label{par:III.A.4.c}

\begin{figure}[t]
\includegraphics[width=8.6cm]{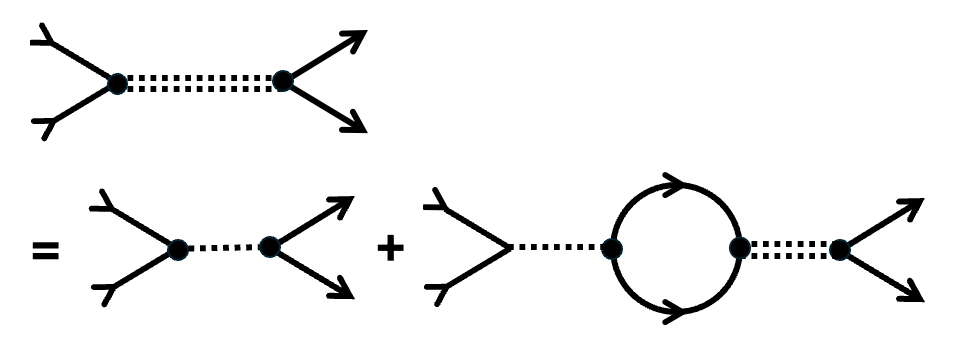}
\caption{The Cooper-screened particle-particle interaction amplitude, denoted by a double dotted line.}
\label{fig:4.1}
\end{figure}

\begin{figure}[t]
\includegraphics[width=8.6cm]{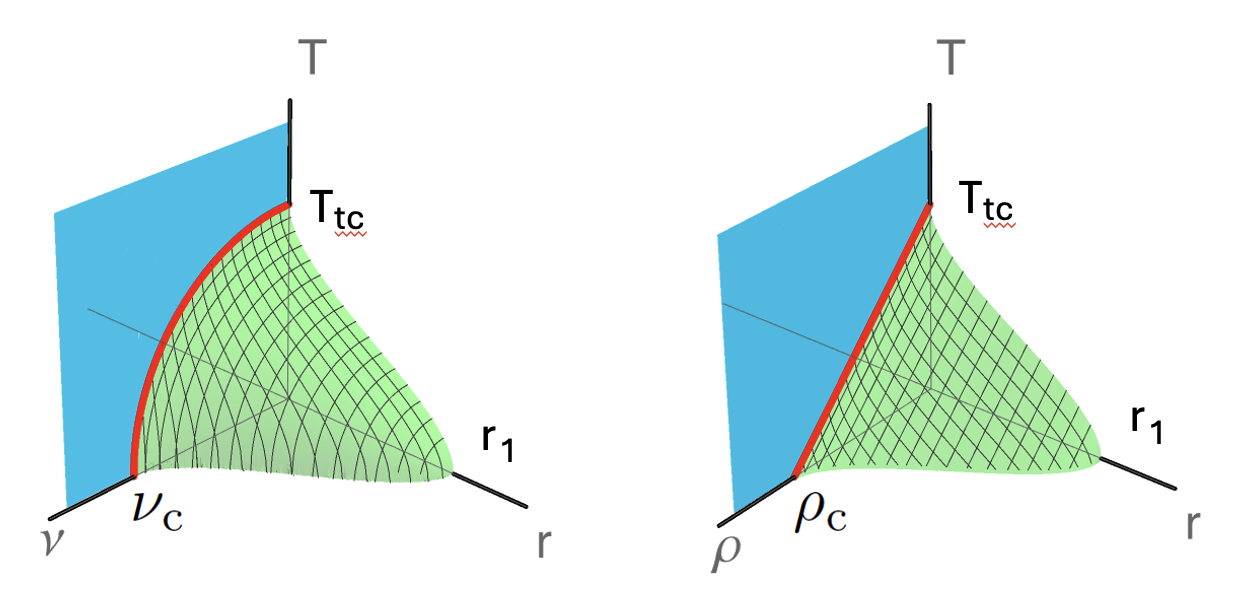}
\caption{Schematic phase diagram in the space spanned by $T$, $r$, and $\nu$, and $T$, $r$, and $\rho$,
              respectively. Shown are surfaces of first-order transition (green, meshed) and second-order
              transitions (blue, solid) that meet at a line of tricritical points (red)}
\label{fig:5}
\end{figure}

As we mentioned in Sec.~\ref{subsubsec:II.A.3}, the effects of the SM1 shown in Eq.~(\ref{eq:3.16})
are weaker than those of the corresponding SM1s in the particle-hole channel. The reason is the Cooper
screening, or screening by particle-particle pairs, of the interaction amplitude in the particle-particle
channel shown in Fig.~\ref{fig:4.1}, which results in the SM1 contribution to the spin susceptibility to be
suppressed by a factor of $1/\ln^2 h$ compared to the corresponding nonanalyticity from a SM1 in the
particle-hole channel. Note that the logarithmic suppression is independent of the dimensionality. 
In 3-d systems, the resulting nonanalyticity in the free energy is $m^4/\ln(1/m)$, which is subleading
to the $m^4$ term in the usual Landau theory.\cite{Belitz_Kirkpatrick_Vojta_1997, Belitz_Kirkpatrick_2026, Miserev_Loss_Klinovaja_2026}
Accordingly, the quantum ferromagnetic transition in 3-d systems
is of second order, provided the coupling constant $\vso$ is sufficiently large. 
The equation of state in 3-d takes the form\cite{Kirkpatrick_Belitz_2020}
\be
h = r\,m - v\,m^3 \ln\left(\frac{1}{m^2 + \nu^2 + (T+\rho)^2}\right) + u\,m^3\ .
\label{eq:3.17}
\ee
Here $\nu = O(\vso/\vF)$ is the dimensionless strength of the spin-orbit interaction, and we have added the effect
of weak quenched disorder as in Eq.~(\ref{eq:3.4}).
The spin-orbit interaction, a nonzero temperature, and weak quenched disorder affect the quantum phase 
transition in qualitatively the same way. The resulting phase diagram in the space spanned by $r$, $T$, 
and the spin-orbit coupling parameter $\nu$ or the disorder parameter $\rho$ is schematically shown in 
Fig.~\ref{fig:5}. The first-order transition is suppressed when $\nu$ or $\rho$ are larger than a critical
value $\nu_{\text{c}}$ and $\rho_{\text{c}}$, respectively. The relevant 
energy scale is the tricritical temperature discussed in Sec.~\ref{subsubsec:III.A.1}, but there is a large 
quantitative renormalization factor involved that is related to the magnetic energy scale, which in all metallic 
magnets is much smaller than the Fermi temperature $\TF$. A semi-quantitative discussion has been given
in Ref.~\onlinecite{Kirkpatrick_Belitz_2020}. The conclusion was that 
a spin-orbit induced 
second order transition is a possibility in ferromagnets such as UIr or CeRh$_6$Ge$_4$. 

In dimensions $d<3$ the nonanalyticity in the free energy is proportional to $m^{d+1}/\ln^2(1/m)$. 
In 2-d systems in particular the equation of state takes the form
\be
h = r\,m - v\,m^2/\ln^2(1/m) + u\,m^3
\label{eq:3.18}
\ee
which leads to a first-order transition.\cite{Miserev_Loss_Klinovaja_2022, Belitz_Kirkpatrick_2026, Miserev_Loss_Klinovaja_2026}

\subsection{Helimagnets}
\label{subsec:III.B}

In helimagnets, a lack of spatial inversion symmetry stabilizes a magnetization pattern that forms a 
global helix with pitch vector ${\bm Q}$:
\be
{\bm m}({\bm x}) = m_0 \left(\cos(Qz), \sin(Qz),0\right)\ ,
\label{eq:3.19}
\ee
where we have chosen ${\bm Q}$ to point in the $z$-direction. The order parameter is the helix amplitude
$m_0$, and the conjugate field is a helically modulated magnetic field
\be
{\bm h}({\bm x}) = h \left(\cos(Qz), \sin(Qz),0\right)\ .
\label{eq:3.20}
\ee

Such a spin structure is realized, e.g., in the metallic compounds MnSi and FeGe as a result of the
Dzyaloshinskii-Moriya interaction that leads to a term ${\bm m}\cdot({\bm\nabla}\times{\bm m})$ in
the classical free-energy functional, which reads\cite{Bak_Jensen_1980, Nakanishi_et_al_1980}
\be
F = \! \int d{\bm x} \left[ \frac{r}{2}\, {\bm m}^2 + \frac{a}{2}\left(\nabla{\bm m}\right)^2 + \frac{c}{2}\,{\bm m}\cdot\left({\bm\nabla}\times{\bm m}\right) + \frac{u}{4}\,{\bm m}^4\right]
\label{eq:3.21}
\ee
The pitch wave number is given by $c/2a$;\cite{Dzyaloshinskii_1958, Moriya_1960} it
has the same value in both the helically ordered ($m_0 \neq 0$) and disordered ($m_0 = 0$) phases. 

The physical origin of the Dzyaloshiskii-Moriya interaction is the spin-orbit interaction, but in contrast
to the spin-orbit terms of Rashba-Dresselhaus type\cite{Dyakonov_2008} in Secs.~\ref{subsubsec:III.A.3} 
and \ref{subsubsec:III.A.4} the Dzyaloshinskii-Moriya interaction is a true electron-electron interaction 
that appears in the action as a four-fermion term and does not modify the single-particle Hamiltonian 
other than via the conjugate field.\cite{DM_footnote} The single-particle Hamiltonian is no longer 
diagonal in wave-vector space and reads
\bea
{\mathcal H}_{{\bm k}{\bm p}} &=& \delta_{{\bm k}{\bm p}}\,\xi_{\bm k}\,\sigma_0 - {\bm h}({\bm k}-{\bm p})\cdot{\bm\sigma}
\nonumber\\
                                               &=& \delta_{{\bm k}{\bm p}}\,\xi_{\bm k}\,\sigma_0 - h\,\delta_{{\bm k},{\bm p}+{\bm Q}}\, \sigma_-
                                                                                                                               - h\,\delta_{{\bm k},{\bm p}-{\bm Q}}\,\sigma_+  \ ,\qquad                                        
\label{eq:3.22}
\eea
where $\sigma_{\pm} = (\sigma_1 \pm i \sigma_2)/2$. The resulting Green function can be expressed in terms of linear 
combinations of the quantities
\bse
\label{eqs:3.23}
\bea
a_{\alpha}({\bm k},{\bm Q};i\omega_m) = \frac{G_0^{-1}({\bm k}+\alpha{\bm Q},i\omega_m)}
     {G_0^{-1}({\bm k},i\omega_m)\,G_0^{-1}({\bm k}+\alpha{\bm Q},i\omega_m) - h^2}
     \nonumber\\
\label{eq:3.23a}\\     
b_{\alpha}({\bm k},{\bm Q};i\omega_m) = \frac{-h}
           {G_0^{-1}({\bm k},i\omega_m)\,G_0^{-1}({\bm k}+\alpha{\bm Q},i\omega_m) - h^2}\ ,
           \nonumber\\
\label{eq:3.23b}           
\eea         
with $\alpha = \pm 1$, as follows,\cite{Belitz_Kirkpatrick_Rosch_2006a}
\bea
\mathcal{G}_{{\bm k}{\bm p}}(i\omega_m) &=& \delta_{{\bm k}{\bm
p}}\,\bigl[\sigma_{+-}\,a_+({\bm k},{\bm q};i\omega_m) + \sigma_{-+}\,a_-({\bm
k},{\bm q};i\omega_m)\bigr]
\nonumber\\
&&\hskip 20pt + \delta_{{\bm k}+{\bm q},{\bm p}}\,\sigma_+\,b_+({\bm k},{\bm
q};i\omega_m)
\nonumber\\
&&\hskip 20pt   + \delta_{{\bm k}-{\bm q},{\bm p}}\,\sigma_-\,b_-({\bm k},{\bm
q};i\omega_m).
\label{eq:3.23c}
\eea
\ese
Here $\sigma_{\pm} = (\sigma_1 \pm i\sigma_2)/2$, $\sigma_{+-} =
\sigma_+\sigma_-$, and $\sigma_{-+} = \sigma_-\sigma_+$.

\begin{figure}[t]
\includegraphics[width=8.5cm]{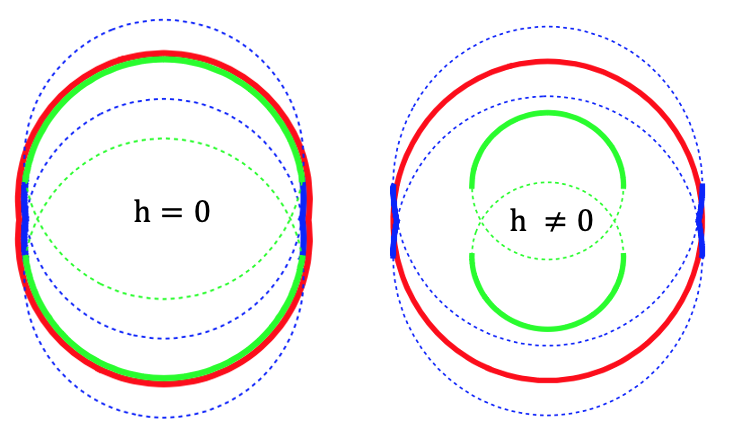}
\caption{Schematic Fermi surface of a helimagnet with $Q \ll 2\kF$. The solid contours denote the Fermi surface 
              associated with the single-particle energy $E_{\bm k}^{++}$ (red) and the parts of the Fermi surfaces 
              associated with $E_{{\bm k}\pm{\bm Q}}^{++}$ (blue) and $E_{{\bm k}\pm{\bm Q}}^{-+}$ (green), respectively, 
              that are degenerate with $E_{\bm k}^{++}$. Parts of $E_{{\bm k}\pm{\bm Q}}^{++}$ and $E_{{\bm k}\pm{\bm Q}}^{-+}$ 
              that are not degenerate with $E_{\bm k}^{++}$ are shown as thin dashed. A nonzero field lifts the degeneracies 
              except in special points; the solid lines refer to the parts that are degenerate in zero field. }
\label{fig:6}
\end{figure}

A partial fraction decomposition of the denominators in these expressions
yields the quasiparticle resonances
\bse
\label{eqs:3.24}
\be
F_k^{\sigma\alpha} = \frac{1}{i\omega_m - \xi_{\bm k} - \nu_{\bm k}(\alpha{\bm Q}) + \sigma w_{\bm k}(\alpha{\bm Q},h)}
\label{eq:3.24a}
\ee
Here
\be
w_{\bm k}({\bm Q},h) =  \sqrt{\left(\nu_{\bm k}({\bm Q})\right)^2 + h^2}
\label{eq:3.24b}
\ee
with
\be
\nu_{\bm k}({\bm Q}) = {\bm k}\cdot{\bm Q}/2m + {\bm Q}^2/4m\ .
\label{eq:3.24c}
\ee
\ese
The Fermi surface consists of four branches labeled by $\alpha$ and $\sigma$, but only two of those are
independent due to the identity
\be
F_{k+Q}^{\sigma,-} = F_k^{\sigma,+}\ ,
\label{eq:3.25}
\ee
where $Q = (i0,{\bm Q})$. This reflects the fact that the Hamiltonian has only two eigenvalues per 
${\bm k}$-value.\cite{counting_footnote} In zero field, two of these branches
are degenerate, and this degeneracy is lifted by a nonzero conjugate field, see Fig.~\ref{fig:6}.

Since the conjugate field, Eq.~(\ref{eq:3.20}), has a helical structure with pitch wave vector ${\bm Q}$,
the relevant susceptibility must be written in terms of convolutions of the form
\be
\frac{1}{V} \sum_{\bm k} F_k^{\sigma\alpha} F_{k\pm Q - q}^{\sigma'\alpha'}
\label{eq:3.26}
\ee
instead of Eq.~(\ref{eq:2.5a}). The appropriate generalization of Eq.~(\ref{eq:2.5c}) then is
\bse
\label{eqs:3.27}
\bea
\phi^{\sigma\sigma',\alpha\alpha'}({\bm q},i\Omega_n) &=& 
\nonumber\\
&& \hskip -75pt
\NF \int\frac{d\Omega_{\bm k}}{4\pi}\ \frac{1}
   {i\Omega_n - \vF\hat{\bm k}\cdot{\bm q} - E_{\bm k}^{\sigma\alpha} + E_{{\bm k}\pm{\bm Q}-{\bm q}}^{\sigma'\alpha'}}\ . 
   \qquad\quad
\label{eq:3.27a}
\eea
with single-particle energies
\be
E_{\bm k}^{\sigma\alpha} = \nu_{\bm k}(\alpha{\bm Q}) - \sigma w_{\bm k}(\alpha{\bm Q},h)\ .
\label{eq:3.27b}
\ee
\ese
To check for the relevant degeneracies, Eq.~(\ref{eq:3.25}) implies that we need to consider only two of the
four single-particle energies; we choose the $E_{\bm k}^{\sigma +}$. An inspection shows that, in zero field,
$E_{\bm k}^{++}$ is degenerate with parts of $E_{{\bm k}\pm{\bm Q}}^{++}$ and parts of $E_{{\bm k}\pm{\bm Q}}^{-+}$,
see Fig.~\ref{fig:6}. A nonzero field lifts these degeneracies except in special points. This gives rise to a soft mode
of the first kind, which is a composite mode with parts of the angular integral in Eq.~(\ref{eq:3.27a}) contributing
for different values of $\sigma$. Note that the
Dzyaloshinskii-Moriya interaction has a qualitatively different effect than the Rashba-Dresselhaus interaction
in Sec.~\ref{subsubsec:III.A.4}, which completely splits the Fermi surface even in zero field, 
Fig.~\ref{fig:4}.\cite{TV_disagreement_footnote}

\begin{figure}[h]
\includegraphics[width=8.5cm]{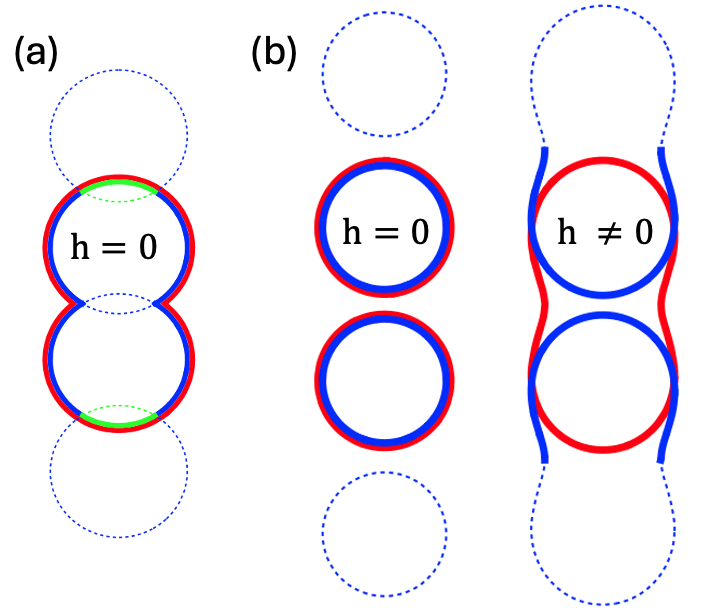}
\caption{Schematic Fermi surface of a helimagnet with $Q \alt 2\kF$ (a) and $Q > 2\kF$ (b). The degeneracies persist and are again
              lifted by a nonzero field except in special points.}
\label{fig:7}
\end{figure}

The sketch in Fig.~\ref{fig:6} applies to the case $Q \ll \kF$. With increasing $Q$
the contributions of  $E_{{\bm k}\pm{\bm Q}}^{++}$ to the soft mode grow and those of $E_{{\bm k}\pm{\bm Q}}^{-+}$
shrink, see panel (a) in Fig.~\ref{fig:7}. For $Q > 2 \kF$ there are six disjoint Fermi surfaces corresponding to $E_{\bm k}^{++}$, 
$E_{{\bm k}+{\bm Q}}^{++}$, and $E_{{\bm k}-{\bm Q}}^{++}$, respectively, four of which are still pairwise
degenerate, see panel (b) in Fig.~\ref{fig:7}. 
A nonzero field again lifts the degeneracy, so there still is a soft mode of the first kind. In this
case our continuum model is physically realistic only if $\kF$ is small compared to the inverse
lattice spacing. Helimagnetism has been observed in semiconductors (for an example, see
Ref.~\onlinecite{Giebultowicz_et_al_1993}), so this is physically realizable. The soft-mode
structure on a lattice requires a separate analysis. For a simple lattice model for spin-density
waves and antiferromagnets, see Sec.~\ref{subsec:III.D} below. 

Even though the overall soft-mode structure is very similar to the case of an ordinary ferromagnet, 
Sec.~\ref{subsubsec:III.A.1}, there is a crucial difference: As can be seen from Eq.~(\ref{eq:3.24b}), 
the field-dependent mass of the SM1 in the helimagnetic case is proportional to $h^2/\vF Q$, as opposed to 
$h$ in the ferromagnet.\cite{HM_mass_footnote} As a result, the QPT problem maps onto the ferromagnetic 
case with $m$ in the nonanalytic term replaced by $m^2/\vF Q$, and the 3-d equation of state takes the form
\be
h = r\,m + u\,m^3 + O(m^5\ln m, m^5)\ .
\label{eq:3.28}
\ee
That is, the QPT is of second order, even though the free energy and the equation of state are nonanalytic
functions of the order parameter. This is because the nonanalyticity is not strong enough to change the order
of the transition. In 2-d the soft-mode contribution competes with the ordinary $u m^3$ term and may
or may not change the order of the transition, see the discussion after Eq.~(\ref{eq:1.12b}).

The above conclusions hold if $\vF Q$ is sufficiently large.
The relevant energy scale is the same as for the strength of the spin-orbit interaction in ferromagnets in
Sec.~\ref{subsubsec:III.A.1}, viz., the tricritical temperature in the absence of an ordering wave number.
One also again needs to take into account the large renormalization factor mentioned after
Eq.~(\ref{eq:3.17}). This is the same renormalization that brings the magnetic energy scale $T_{\text m}$ of a ferromagnet
down from the atomic energy scale $\TF$; $\TF/T_{\text{m}}$ is on the order of several hundred in low-temperature magnets.
The conclusion is that the transition is second order if $\vF Q$ is larger than the renormalized tricritical temperature
by a factor of about $\TF/T_{\text{m}}$.
For $\vF Q$ small compared to that scale the problem crosses
over to the ordinary ferromagnetic one, the equation of state is given by Eqs.~(\ref{eqs:3.3}), and the
transition is of first order. The estimates in Ref.~\onlinecite{Kirkpatrick_Belitz_2020} indicate that
this is the situation in the long-wavelength helimagnet MnSi.

\subsection{Magnetic Nematics}
\label{subsec:III.C}

In a magnetic nematic the order parameter is a tensor-valued modulated spin density
of the form\cite{Oganesyan_Kivelson_Fradkin_2001, Wu_et_al_2007}  
\be
N_i = \langle\bar\psi_a(x)\,\sigma_i^{ab} f({\bm\nabla}_{\bm x})\,\psi_b(x)\rangle\ ,
\label{eq:3.29}
\ee
with $f$ a tensor-valued monomial function of the gradient operator. 
In the simplest case, i.e., a p-wave nematic, $f$ is a linear function of the gradient, 
$f({\bm\nabla}_{\bm x}) = i{\bm\nabla}_{\bm x}$, and the order parameter $N_i^{\alpha}$
carries a spin index $i$ and one orbital index $\alpha$. It is obvious how to generalize
to higher rank tensors. A homogeneous conjugate field can formally be constructed as follows.
Let ${\bm h}({\bm x})$ be a magnetic field that is a linear function of the position,
$h_i({\bm x}) = {\tilde h}_i^{\alpha} x_{\alpha}$. Then the field conjugate to $N_i^{\alpha}$ is
given by\cite{Kirkpatrick_Belitz_2011}
\be
h_N^{i\,\alpha} = \epsilon_{ijk}\,\frac{1}{V}\int d{\bm y}\,h_j({\bm y})\,{\tilde h}_k^{\alpha}\ .
\label{eq:3.30}
\ee

\begin{figure}[t]
\includegraphics[width=8.5cm]{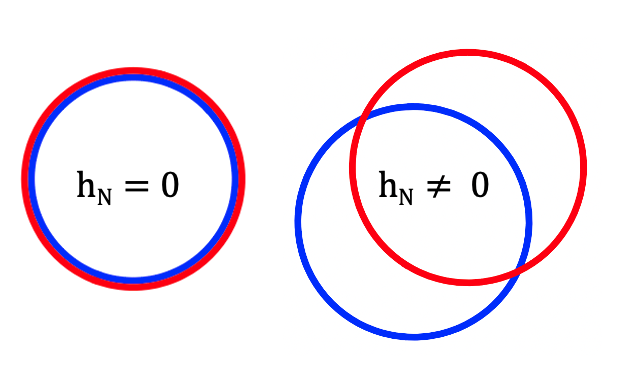}
\caption{Schematic Fermi surface of a p-wave magnetic nematic. The two degenerate sheets are split by the conjugate
              field.}
\label{fig:8}
\end{figure}

The nematic, or non-$s$-wave, magnetic order is caused by electron-electron interactions.
The single-particle Hamiltonian is not affected and reads
\bse
\label{eqs:3.31}
\be
{\mathcal H}_{\bm k} = \xi_{\bm k} \sigma_0 - {\frak h}_{\bm k}\cdot{\bm\sigma}
\label{eq:3.31a}
\ee
where
\be
{\frak h}_{\bm k} = h_N^{i\,\alpha} k_{\alpha}\ .
\label{eq:3.31b}
\ee
\ese
We see that the soft-mode problem maps onto the ferromagnetic case with ${\bm\sigma}\cdot{\bm h}$
replaced by ${\frak h}_{\bm k}\cdot{\bm\sigma} = \sigma_i\,h_N^{i\,\alpha} k_{\alpha}$; the field-split
Fermi surface is shown in Fig.~\ref{fig:8}. Accordingly, the soft modes analogous to Eq.~(\ref{eq:2.5c}) 
are given by
%
\bea
\phi^{\sigma\sigma'}({\bm q},i\Omega_n) &=& 
\nonumber\\
&&\hskip -70pt
\NF \int \frac{d\Omega_{\bm k}}{4\pi}\,\frac{1}{i\Omega_n - \vF\hat{\bm k}\cdot{\bm q}
      - \sigma'\,\hat{\frak h}_{\bm k}^i h_N^{i\,\alpha} q_{\alpha} - (\sigma - \sigma')\vert{\frak h}_{\bm k}\vert}\ .
      \nonumber\\
\label{eq:3.32}
\eea
We see that the modes with $\sigma \neq \sigma'$ are soft of the first kind with respect to the
conjugate field. Accordingly, the quantum phase transition maps onto the ferromagnetic one and
is of first order. In $d=3$ the equation of state
is given by Eq.~(\ref{eq:3.4}) with $m$ replaced by the nematic order parameter, as was concluded in
Ref.~\onlinecite{Kirkpatrick_Belitz_2011} by explicitly constructing the equation of state. The same
conclusion holds for magnetic nematics with higher angular momentum, i.e., with the operator
$f({\bm\nabla}_{\bm x})$ in Eq.~(\ref{eq:3.29}) a higher-rank tensor.

\subsection{Magnetic Smectics, and N{\'e}el Antiferromagnets}
\label{subsec:III.D}

We next consider magnetic smectics, i.e., a unidirectional spin-density wave or spin-stripe system characterized by
a modulated magnetization given by
\be
{\bm m}({\bm x}) = m_0 \left(\cos(Qz), 0, 0\right)\ .
\label{eq:3.33}
\ee
As in the helimagnetic case we have chosen ${\bm Q}$ to point in the $z$-direction. The order parameter is 
given by the amplitude $m_0$, the conjugate field is
\be
{\bm h}({\bm x}) = h \left(\cos(Qz), 0, 0\right)\ ,
\label{eq:3.34}
\ee
and the single-particle Hamiltonian is given by
\be
{\mathcal H}_{{\bm k}{\bm p}} = \delta_{{\bm k}{\bm p}}\,\xi_{\bm k}\,\sigma_0 
     - \frac{1}{2}\,h \left(\delta_{{\bm k},{\bm p}+{\bm Q}} + \delta_{{\bm k},{\bm p}-{\bm Q}}\right) \sigma_1\ .
\label{eq:3.35}
\ee

Note that the smectic can be considered a superposition of two helices with pitch wave vectors ${\bm Q}$
and $-{\bm Q}$, respectively, see Eq.~(\ref{eq:3.19}). Despite this fact, the quasiparticle resonance spectrum
is much harder to determine than in the helimagnetic case, and we have been unable to do so exactly.
We therefore resort to an approximate scheme that suffices for our purposes. In order to do so, it is
illustrative to first apply this method to the helical case. 

\subsubsection{Helimagnets revisited}
\label{subsubsec:III.D.1}

Consider the helimagnetic Hamiltonian from Eq.~(\ref{eq:3.22}) and the corresponding eigenproblem,
Eq.~(\ref{eq:A.1}). Writing the eigenvector as ${\bm v}_{\bm k} = (u_{\bm k},w_{\bm k})$,
we have
\bse
\label{eqs:3.36}
\bea
(\lambda - \xi_{\bm k}) u_{\bm k} &=& -h\,w_{{\bm k}+{\bm Q}}\ ,
\label{eq:3.36a}\\
(\lambda - \xi_{\bm k}) w_{\bm k} &=& -h\,u_{{\bm k}-{\bm Q}}\ .
\label{eq:3.36b}
\eea
\ese
For $h=0$, this yields one doubly degenerate eigenvalue per ${\bm k}$ vector, $\lambda = \xi_{\bm k}$. 
While this is consistent with the fact that the Hamiltonian at $h=0$ is the free-fermion Hamiltonian, it
is misleading for the purpose of finding the quasiparticle resonances. To see this, we eliminate $w$ from
Eq.~(\ref{eq:3.36a}) and $u$ from Eq.~(\ref{eq:3.36b}) to find
\bse
\label{eqs:3.37}
\bea
\left[\left(\lambda - \xi_{{\bm k}+{\bm Q}}\right)\left(\lambda - \xi_{\bm k}\right) - h^2\right] u_{\bm k} &=& 0\ ,
\label{eq:3.37a}\\
\left[\left(\lambda - \xi_{{\bm k}-{\bm Q}}\right)\left(\lambda - \xi_{\bm k}\right) - h^2\right] w_{\bm k} &=& 0\ .
\label{eq:3.37b}
\eea
\ese
These equations yield the four values for $\lambda$ given in Eqs.~(\ref{eqs:A.2}). In particular, for $h=0$
we have $\lambda = \xi_{{\bm k}\pm{\bm Q}}$ in addition to the doubly degenerate $\lambda = \xi_{\bm k}$.
If one interprets these four $\lambda$ values as eigenvalues of the Hamiltonian, only two are independent,
see the discussion in Appendix~\ref{app:A}. However, all four of them appear in the denominator of the
Green function, see Eqs.~(\ref{eqs:3.23}), and they therefore represent independent quasiparticle resonances. 
The reason is the singular nature, as $h\to 0$, of the parts of the Green function that are not diagonal in
wave-vector space, see Eq.~(\ref{eq:3.23b}), which have the structure of a delta function. 

Now consider the four single-particle energies given by the four values of $\lambda$. At $h=0$ they are
given by $\xi_{\bm k}$, $\xi_{{\bm k}\pm{\bm Q}}$, and their union is identical with the union of the four
branches of $E_{\bm k}^{\sigma\alpha}$ in Eq.~(\ref{eq:3.27b}). For small $h\neq 0$ we obtain corrections
of $O(h^2)$,
\bse
\label{eqs:3.38}
\bea
E^{(1)}_{\bm k} &=& \xi_{\bm k} + h^2/(\xi_{\bm k}-\xi_{{\bm k}+{\bm Q}}) + O(h^4)\ ,
\label{eq:3.38a}\\
E^{(2)}_{\bm k} &=& \xi_{{\bm k}+{\bm Q}} - h^2/(\xi_{\bm k}-\xi_{{\bm k}+{\bm Q}}) + O(h^4)\ ,\quad
\label{eq:3.38b}\\
E^{(3)}_{\bm k} &=& E^{(2)}_{{\bm k}-{\bm Q}}\ ,
\label{eq:3.38c}\\
E^{(4)}_{\bm k} &=& E^{(1)}_{{\bm k}-{\bm Q}}\ .
\label{eq:3.38d}
\eea
\ese
We see that in zero field $E_{{\bm k}+{\bm Q}}^{(1)}$ and $E_{\bm k}^{(2)}$ are degenerate, and
so are  $E_{{\bm k}-{\bm Q}}^{(3)}$ and $E_{\bm k}^{(4)}$. The expansion in powers of $h^2$
breaks down in parts of momentum space, but it suffices to demonstrate that a nonzero field
lifts these degeneracies. 

All of this is consistent with the results in Sec.~\ref{subsec:III.B} which did not rely on an expansion in 
powers of the field. The method used above, which is simpler but less complete than
the closed-form solution in Sec.~\ref{subsec:III.B}, also works for the spin-wave case, as we will now
demonstrate. The quantitative considerations regarding the size of $\vF Q$ from the end of
Sec.~\ref{subsec:III.B} apply equally to helimagnets, magnetic smectics, and antiferromagnets.

\subsubsection{Magnetic smectics}
\label{subsubsec:III.D.2}

Now consider the spin-wave Hamiltonian, Eq.~(\ref{eq:3.35}). The eigenequations analogous to
Eqs.~(\ref{eqs:3.36}) read
\bse
\label{eqs:3.39}
\bea
(\lambda - \xi_{\bm k}) u_{\bm k} &=& \frac{-h}{2}\left(w_{{\bm k}+{\bm Q}} + w_{{\bm k}-{\bm Q}}\right)\ ,
\label{eq:3.39a}\\
(\lambda - \xi_{\bm k}) w_{\bm k} &=& \frac{-h}{2}\left(u_{{\bm k}+{\bm Q}} + u_{{\bm k}-{\bm Q}}\right)\ .
\label{eq:3.39b}
\eea
\ese
We now follow the procedure from Sec.~\ref{subsubsec:III.D.1}, i.e., we
use Eq.~(\ref{eq:3.39b}) to eliminate $w$ from Eq.~(\ref{eq:3.39a}) and Eq.~(\ref{eq:3.39a}) to eliminate
$u$ from Eq.~(\ref{eq:3.39b}). We then obtain identical equations for $u$ and $w$,
\bea
\left[\lambda - \xi_{\bm k} - \frac{h^2}{4}\left(\frac{1}{\lambda-\xi_{{\bm k}+{\bm Q}}} + \frac{1}{\lambda-\xi_{{\bm k}-{\bm Q}}} 
   \right)\right] x_{\bm k} &=&
\nonumber\\
&& \hskip -200pt \frac{h^2}{4} \left( \frac{1}{\lambda - \xi_{{\bm k}+{\bm Q}}}\,x_{{\bm k}+2{\bm Q}}
                                              +  \frac{1}{\lambda - \xi_{{\bm k}-{\bm Q}}}\,x_{{\bm k}-2{\bm Q}}\right) \ , \qquad\quad
\label{eq:3.40}
\eea
where $x$ can stand for either $u$ or $w$. In contrast to the helical case, $x_{\bm k}$ couples to $x_{{\bm k}\pm 2{\bm Q}}$.
Iterating the equation, we see that couplings to higher multiples of ${\bm Q}$ are accompanied by higher
powers of $h$. Iterating once, and truncating at that stage, we obtain
\bse
\label{eqs:3.41}
\bea
\Bigl[(\lambda &-& \xi_{\bm k})(\lambda - \xi_{{\bm k}+{\bm Q}})(\lambda -\xi_{{\bm k}-{\bm Q}}) 
\nonumber\\
&& \hskip -20pt - \frac{h^2}{4} (2\lambda - \xi_{{\bm k}-{\bm Q}} - \xi_{{\bm k}+{\bm Q}}) + O(h^4)\Bigr] x_{\bm k} 
   = O(h^4 x_{{\bm k}\pm 4{\bm Q}})\ ,
   \nonumber\\
\label{eq:3.41a}
\eea
which yields two identical cubic equations
\bea
(\lambda - \xi_{\bm k})(\lambda - \xi_{{\bm k}+{\bm Q}})(\lambda -\xi_{{\bm k}-{\bm Q}}) = && \hskip -10pt
    \frac{h^2}{4} (2\lambda - \xi_{{\bm k}-{\bm Q}} - \xi_{{\bm k}+{\bm Q}})
    \nonumber\\
    &&+ O(h^4)
\label{eq:3.41b}
\eea
\ese
instead of the two quadratic ones from Eqs.~(\ref{eqs:3.37}). 
To this order in the expansion in powers of $h^2$ this suggests three single-particle energy branches that all are
two-fold degenerate. At $h=0$, they are given by $\xi_{\bm k}$, $\xi_{{\bm k}+{\bm Q}}$, and $\xi_{{\bm k}-{\bm Q}}$, 
respectively. For small $h\neq 0$ they are
\bse
\label{eqs:3.42}
\bea
E_{\bm k}^{(1)} &=& \xi_{\bm k} + h^2\,\frac{2\xi_{\bm k} - \xi_{{\bm k}-{\bm Q}} - \xi_{{\bm k}+{\bm Q}}}{4(\xi_{\bm k}-\xi_{{\bm k}+{\bm Q}})(\xi_{\bm k}-\xi_{{\bm k}-{\bm Q}})}\ ,\qquad
\label{eq:3.42a}\\
E_{\bm k}^{(2)} &=& \xi_{{\bm k}+{\bm Q}} + h^2\,\frac{1}{4(\xi_{{\bm k}+{\bm Q}}-\xi_{\bm k})}\ ,
\label{eq:3.42b}\\
E_{\bm k}^{(3)} &=& \xi_{{\bm k}-{\bm Q}} + h^2\,\frac{1}{4(\xi_{{\bm k}-{\bm Q}}-\xi_{\bm k})}\ .
\label{eq:3.42c}
\eea
\ese
For a proper interpretation of this result we need to ascertain that the $E_{\bm k}^{(i)}$ do indeed appear as poles in
the Green function, as did the $E_{\bm k}^{\sigma\alpha}$ in the helimagnetic case. A perturbative analysis
of the Green function is consistent with this being the case, see Appendix~\ref{app:B}.

As in the helimagnetic case, the relevant susceptibility needs to be taken at an external wave vector equal to $\pm{\bm Q}$,
and the generalization of Eq.~(\ref{eq:2.5c}) is
\bea
\phi^{ij}({\bm q},i\Omega_n) &=& 
\nonumber\\
&& \hskip -75pt
\NF \int\frac{d\Omega_{\bm k}}{4\pi}\ \frac{1}
   {i\Omega_n - \vF\hat{\bm k}\cdot{\bm q} - E_{\bm k}^{(i)} + E_{{\bm k}\pm{\bm Q}-{\bm q}}^{(j)}}\ . 
   \qquad\quad
\label{eq:3.43}
\eea
From Eqs.~(\ref{eqs:3.42}) we see that $E^{(1)}_{{\bm k}+{\bm Q}}$ and  $E^{(2)}_{\bm k}$ are degenerate for $h=0$, 
and so are $E^{(3)}_{{\bm k}+{\bm Q}}$ and  $E^{(1)}_{\bm k}$. A nonzero field lifts these degeneracies. As a result,
there are soft modes of the first kind given by $\phi^{12}$ and $\phi^{13}$. As in the helimagnetic case, the masses
of these soft modes are proportional to $h^2$, and the resulting nonanalyticities are not strong enough to change
the second-order nature of the quantum phase transition in 3-d systems. For $d=2$ the remarks after Eqs.~(\ref{eq:1.12b})
and (\ref{eq:3.28}) apply. 

\begin{figure}[t]
\includegraphics[width=8.5cm]{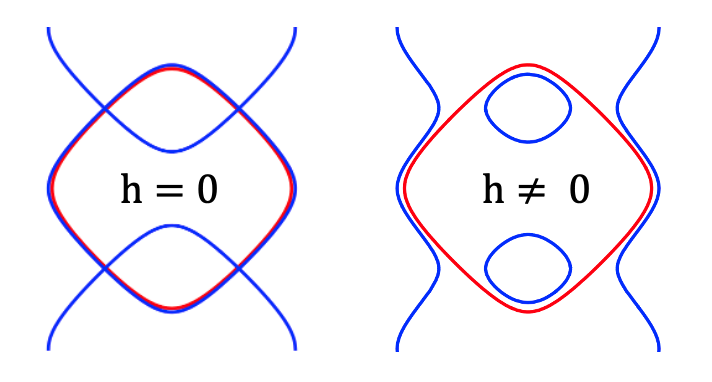}
\caption{Schematic Fermi surface of a magnetic smectic with $Q=\pi/a$. See the text for additional information.}
\label{fig:9}
\end{figure}

In order to find the single-particle energies in the ${\bm k}$-space region where the expansion in Eqs.~(\ref{eqs:3.42})
breaks down one has to discuss the cubic equation in Eq.~(\ref{eq:3.41b}) in detail. Rather then doing so,
we use a simplification that occurs in the case of large $Q$ on the order of a lattice spacing. To see this, we adopt
a lattice model rather than the continuum model we have worked with so far. Within a tight-binding model on a
cubic lattice the single-electron energy is, instead of Eq.~(\ref{eq:2.1b}), 
\be
\epsilon_{\bm k} = -2t\left[\cos(k_x a) + \cos(k_y a) + \cos(k_z a)\right]\ ,
\label{eq:3.44}
\ee
with $t$ the hopping integral. For an ordering wave vector
\be
{\bm Q} = (\pi/a) (0,0,1)
\label{eq:3.45}
\ee
we then have
\be
\xi_{{\bm k}+{\bm Q}} = \xi_{{\bm k}-{\bm Q}} \equiv \xi_{{\bm k}\pm{\bm Q}} = -\epsilon_{\bm k} - \mu
\label{eq:3.46}
\ee
and the cubic equation (\ref{eq:3.41b}) decomposes into a linear equation that is
independent of $h$ and a quadratic one that has the same structure as the quadratic
equation in the helimagnetic case. Accordingly, the single-particle energies are
\bse
\label{eqs:3.47}
\bea
E_{\bm k}^{(1)} &=& \xi_{{\bm k} \pm{\bm Q}}\ ,
\label{eq:3.47a}\\
E_{\bm k}^{(2)} &=& \frac{1}{2} \left[\xi_{\bm k} + \xi_{{\bm k}\pm{\bm Q}} + \sqrt{\left(\xi_{\bm k} - \xi_{{\bm k}\pm{\bm Q}}\right)^2 + 2h^2}\right]\ ,
\nonumber\\
\label{eq:3.47b}\\
E_{\bm k}^{(3)} &=&  \frac{1}{2} \left[\xi_{\bm k} + \xi_{{\bm k}\pm{\bm Q}} - \sqrt{\left(\xi_{\bm k} - \xi_{{\bm k}\pm{\bm Q}}\right)^2 + 2h^2}\right]\ .
\nonumber\\
\label{eq:3.47c}
\eea
\ese
We note that within our simple tight-binding model and the approximations that led to Eq.~(\ref{eq:3.41b})
the energy branches $E_{\bm k}^{(2)}$ and $E_{\bm k}^{(3)}$ do not
lead to a Fermi surface at half filling, $\mu=0$,  if $h\neq 0$. This case requires a more complete analysis and we will restrict ourselves to a
system away from half filling in what follows. In zero field, we can relabel the energy branches such that 
the second and third one become
\bse
\label{eqs:3.48}
\bea
{\tilde E}_{\bm k}^{(2)} &=& \xi_{\bm k}\ ,
\label{eq:3.48a}\\
{\tilde E}_{\bm k}^{(3)} &=& \xi_{{\bm k}\pm{\bm Q}}\ .
\label{eq:3.48b}
\eea
\ese
We see that $E_{{\bm k}\pm{\bm Q}}^{(1)}$ is degenerate with ${\tilde E}_{\bm k}^{(2)}$, and this degeneracy
is lifted by a nonzero field, see Fig.~\ref{fig:9}. Accordingly, there is a soft mode of the first kind that is
qualitatively similar to the helimagnetic case, see Sec.~\ref{subsec:III.B}, and the equation of state is
given by Eq.~(\ref{eq:3.28}). This demonstrates that the conclusions drawn from Eqs.~(\ref{eqs:3.42})
are qualitatively correct and remain valid in the framework of a lattice model. 

The analysis presented above can obviously be continued to higher order in powers of $h^2$ by iterating
Eq.~(\ref{eq:3.40}). This yields successively higher-order polynomials for the $\lambda$ and hence the
single-particle energies. It is easy to see that, in zero field, there are resonances at $\xi_{{\bm k}+n{\bm Q}}$
where $n$ can be any integer. While this produces additional soft modes, it does not affect the existence
of the soft modes discussed above.

\subsubsection{N{\'e}el Antiferromagnets}
\label{subsec:III.D.3}

\begin{figure}[t]
\includegraphics[width=8.5cm]{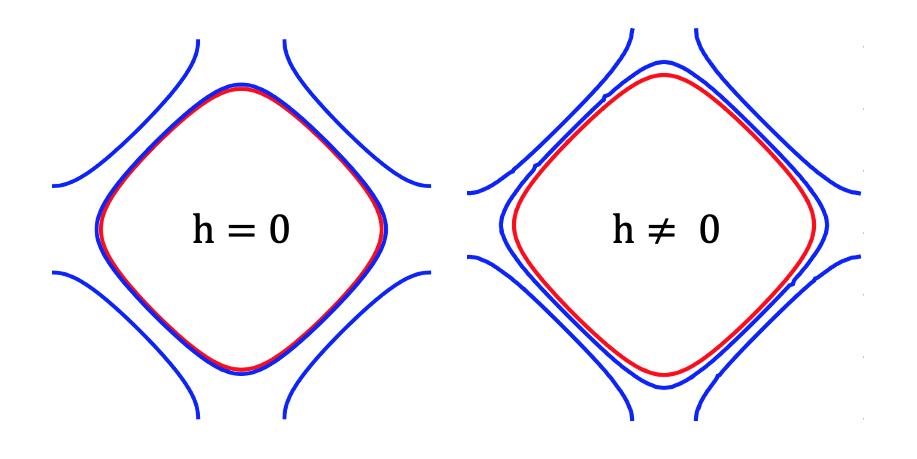}
\caption{Schematic Fermi surface of a N{\'e}el AFM. See the text for additional information.}
\label{fig:10}
\end{figure}

In N{\'e}el antiferromagnets, i.e., collinear antiferromagnets with fully compensated magnetization, the 
order parameter is the staggered magnetization, which can be modeled as
\bse
\label{eqs:3.49}
\be
{\bm N}({\bm x}) = N_0 \left(\cos ( {\bm Q}\cdot{\bm x}),0,0\right)\ ,
\label{eq:3.49a}
\ee
with an ordering wave vector 
\be
{\bm Q} = (\pi/a)(1,1,1)\ ,
\label{eq:3.49b}
\ee
\ese
with $a$ the lattice spacing. We have chosen the staggered magnetization
to point in the $x$-direction. The conjugate field is a staggered magnetic field
\be
{\bm h}({\bm x}) = h\left(\cos ({\bm Q}\cdot{\bm x}),0,0\right)\ ,
\label{eq:3.50}
\ee
and the Hamiltonian is given by Eq.~(\ref{eq:3.35}) with ${\bm Q}$ from Eq.~(\ref{eq:3.49b}).
We again adopt the tight-binding model, Eq.~(\ref{eq:3.44}). Since the Hamiltonian is the
same as in the smectic case, except for the different ${\bm Q}$ vector, the analysis of the
soft-mode structure is exactly analogous to the latter, and so is the conclusion. The 
schematic Fermi surfaces away from half filling are shown in Fig.~\ref{fig:10}.
There is a soft mode of the first kind leading to a nonanalytic equation of state, but the
nonanalyticity is not strong enough to modify the order of the transition, which is second
order with the equation of state given by Eq.~(\ref{eq:3.28}).

\subsection{Altermagnets}
\label{subsec:III.E}

\begin{figure}[t]
\includegraphics[width=8.5cm]{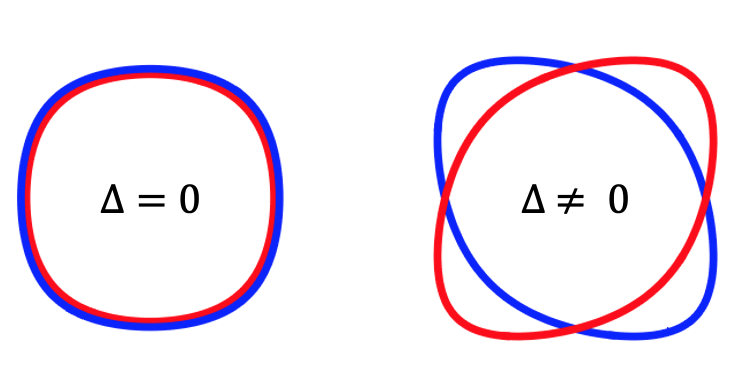}
\caption{Schematic Fermi surface of an altermagnet.}
\label{fig:11}
\end{figure}

Altermagnets and related systems\cite{Hayami_Yanagi_Kusunosel_2019, Smejkal_et_al_2020, Yuan_et_al_2020, Mazin_et_al_2021, Ma_et_al_2021, Hu_et_al_2025, Smejkal_Sinova_Jungwirth_2022, Yuan_Georgescu_Rondinelli_2024, Song_et_al_2025, Mazin_2023, Krempasky_et_al_2024, Fedchenko_et_al_2024, Duan_et_al_2025, Gu_et_al_2025, Smejkal_2024} 
are characterized by an antiferromagnetic order accompanied by a distortion of the Fermi
surface similar to what occurs in a magnetic nematic. However, in contrast to a magnetic nematic, the
Fermi-surface distortion is due to the lattice structure, rather than to electron-electron interaction effects.
In an altermagnet, both types of magnetic order appear simultaneously at the phase transition and have
a common order parameter $\Phi$. 
Neither the antiferromagnetic nor the nematic part of the altermagnetic order results in a homogeneous
magnetization, but the distorted Fermi surface leads to a field-split Fermi surface. A simple 2-d tight-binding
Hamiltonian that reflects these effects and is representative of features of the band structure of RuO$_2$
is \cite{Smejkal_Sinova_Jungwirth_2022b, alternative_Hamiltonians_footnote}
\be
{\mathcal H}_{\bm k} = 2t \cos(k_x a) \cos(k_y a) \sigma_0 + \Delta \sin(k_x a) \sin(k_y a) \sigma_3 - \mu\ .
\label{eq:3.51}
\ee
Here $\Delta$ is proportional to the part of the order parameter that describes the
Fermi surface distortion in the altermagnetic phase, or to the corresponding conjugate field in the
nonmagnetic phase. For $\Delta = 0$ the Fermi surface is two-fold degenerate; a nonzero $\Delta$
splits this degeneracy except in exceptional points, see Fig.~\ref{fig:11}, and the splitting is linear in $\Delta$. 

A spin-orbit interaction
could lead to a split Fermi surface even in zero field, see Sec.~\ref{subsubsec:III.A.4}. However, in
these systems the spin-orbit interaction is weak, so there is a soft mode of the first kind. All of this is 
reflected in Table~\ref{table:I}. According to the 
arguments given in Sec.~\ref{sec:II}, the free energy therefore will be a nonanalytic function of $\Phi$. 
The equation of state is the same as for a magnetic nematic, which in turn maps onto the ferromagnetic
case, see Sec.~\ref{subsec:III.C}. In 2-d it has the form
\be
h_{\Delta} = r\,\Delta -v\,\Delta^2 + u\,\Delta^3\ .
\label{eq:3.52}
\ee
Arguments analogous to those given in Ref.~\onlinecite{Brando_et_al_2016a} show that the sign of
the nonanalytic term is thus that it leads to a first-order QPT (as is the case for ferromagnets and
for magnetic nematics), i.e., $v>0$. That is, as far as the soft modes and their influence on the quantum phase transition
is concerned, altermagnets behave qualitatively the same as ordinary ferromagnets or magnetic
nematics. The complete phase diagram is as shown in Fig.~\ref{fig:5}.

\section{Summary, and Discussion}
\label{sec:IV}

\subsection{Summary}
\label{subsec:IV.A}

To summarize, we have given a classification of all known types of magnetism in metals with respect
to the nature of the magnetic quantum phase transition. One of our key results is that fermionic soft modes lead to long-ranged
order-parameter correlations that are the same in all types of magnets. In clean systems they decay 
in real space as $1/r^{2d-1}$, with $d$ the spatial dimensionality. In wave-number space this corresponds
to a $q^{d-1}$ nonanalyticity with a possible logarithmic factor depending on the dimensionality and
the nature of the soft mode. In clean systems with a homogeneous order parameter (not necessarily
a homogeneous magnetization) the wave number scales as the order parameter. This leads to a
nonanalytic term in the Landau free energy that has a definite sign, which in turn leads to a quantum
phase transition that is of first order in almost all systems. The only exception are 3-d non-centrosymmetric
ferromagnets with a strong spin-orbit interaction, where the nature of the soft mode leads to a
nonanalyticity that is subleading compared to the usual quartic term in the Landau free energy. 
If the magnetic order comes with a nonzero ordering wave vector, such as in helimagnets or
antiferromagnets, the wave number scales as the order parameter squared. As a result, the 
nonanalyticity is not strong enough to modify the order of the quantum phase transition
(provided the ordering wave number is sufficiently large). However, the fermion-induced
long-range correlations are still present and will in general influence the critical behavior. 
All of this is summarized in Table~\ref{table:I}.

In clean 2-d systems with a homogeneous order parameter the transition is also of first order.
For magnets with nonzero ordering wave vector the soft-mode generated term in the free
energy competes with the usual quartic term and no definitive statement about the order
of the transition can be made. 

The presence of quenched disorder leads to even stronger correlations that decay as $1/r^{2d-2}$,
but the sign of the resulting nonanalytic contribution to the free energy is opposite from the one
in the clean case. As a result, the quantum phase transition in 3-d systems is generically of second order,
but with unusual critical behavior. 2-d systems with quenched disorder are more complicated
due to electron localization effects.

\subsection{Discussion}
\label{subsec:IV.B}

In the remainder of this section we discuss various points that not, or only briefly, mentioned in 
the main text.

\subsubsection{Long-range correlations in response and correlation functions}
\label{subsubsec:IV.B.1}

A remarkable aspect of our results is that long-range order-parameter correlations exist in {\em all} known magnets
and are all equally strong. In particular, they are equally strong in magnets with a nonzero ordering wave vector,
such as antiferromagnets, or helimagnets, as in ferromagnets with a homogeneous order parameter. They are
reflected by an order-parameter susceptibility that decays as $1/r^{2d-1}$ in real space, with a multiplicative
logarithm in $d=3$, see Eqs.~(\ref{eq:1.9b}) and (\ref{eq:2.10}). The reason why these correlations change
the order of the quantum phase transition from second to first order in some classes of magnets, but not in
others, is that the order parameter (or, equivalently, the conjugate field) scales differently with the wave number 
in different classes, see Eqs.~(\ref{eq:1.3}) and (\ref{eq:1.4}). 

The order-parameter susceptibility is a response function. In the limit of zero wave number it is given by a
thermodynamic derivative. As was mentioned in Sec.~\ref{subsubsec:II.A.2}, an electron-electron interaction
is necessary for the susceptibility to show the long-range behavior. In addition, the soft-mode spectrum must
be different in the absence and presence, respectively, of a conjugate field for the long-range behavior to
emerge. This explains, for instance, why the spin susceptibility of a Fermi liquid displays long-range behavior,
but the density susceptibility does not.\cite{Belitz_Kirkpatrick_Vojta_1997} It is interesting to compare these
properties of susceptibilities with the corresponding equal-time correlation functions. The two are related by
the fluctuation-dissipation theorem as follows.\cite{Forster_1975} Let $\chi''(q,\omega)$ be the spectral
density of a susceptibility, and let $S(q,\omega)$ be the corresponding correlation function. Then
\be
S(q,\omega) = \frac{2}{1-e^{-\omega/T}}\,\chi''(q,\omega)\ ,
\label{eq:4.1}
\ee
The equal-time correlation function is
\bse
\label{eqs:4.2}
\be
S(q) = \int_{-\infty}^{\infty} \frac{d\omega}{2\pi}\,S(q,\omega) 
        = \int_{-\infty}^{\infty} \frac{d\omega}{\pi}\,\frac{1}{1-e^{-\omega/T}}\,\chi''(q,\omega)\ .
\label{eq:4.2a}
\ee
and the static response function is
\be
\chi(q) = \int_{-\infty}^{\infty} \frac{d\omega}{\pi}\,\chi''(q,\omega)/\omega\ .
\label{eq:4.2b}
\ee
\ese
For wave numbers and frequencies that are small on the microscopic (i.e., atomic) scale we
expect $\chi''$ to be a function of $\omega/\vF q$, 
\be
\chi''(q,\omega) = X(\omega/\vF q)\ , 
\label{eq:4.3}
\ee
with $\vF$ the microscopic velocity scale. Note that this scaling property of the dynamic response
function is very different from Eq.~(\ref{eq:1.1}) for the static order-parameter susceptibility, 
although it has the same origin: susceptibilities are determined by the same type of integrals
that occur in Eqs.~(\ref{eqs:2.5}). The equal-time correlation function then becomes
\bse
\label{eqs:4.4}
\be
S(q) = \vF q\,\Sigma(T/\vF q)\ ,
\label{eq:4.4a}
\ee
with
\be
\Sigma(\tau) = \int_{-\infty}^{\infty} \frac{dx}{\pi}\,\frac{1}{1 - e^{-x/\tau}}\,X(x)\ .
\label{eq:4.4b}
\ee
\ese
Generically we therefore expect $S(q)$ to a linear function of $q$ at $T=0$, corresponding
to a $1/r^{d+1}$ behavior in real space. At small nonzero $T$ it will be a constant 
proportional to $T$ at $\vF q \ll T$ and cross over to a linear $q$-dependence for
larger $q$.

\paragraph{Density-density correlation function}
\label{par:IV.B.1a}

To illustrate these points we choose a very simple example, namely, the density fluctuations 
in a noninteracting Fermi gas.\cite{Pines_Nozieres_1989, Fetter_Walecka_1971} 
The response function $\chi''_{nn}$ has the form (\ref{eq:4.3}) with 
\be
X_{nn}(x) = \frac{\partial n}{\partial\mu}\,\frac{\pi}{2}\,x\,\Theta(1-\vert x\vert)
\label{eq:4.5}
\ee
and the scaling function $\Sigma$ in Eqs.~(\ref{eqs:4.4}) is
\bse
\label{eqs:4.6}
\be
\Sigma_{nn}(\tau) = \frac{\partial n}{\partial\mu}\,\frac{1}{2} \int_0^1 dx\,x\,\coth(x/2\tau)\ .
\label{eq:4.6a}
\ee
The asymptotic behavior for small and large $\tau$, respectively, is
\be
\Sigma_{nn}(\tau) = \frac{\partial n}{\partial\mu} \times 
                                \begin{cases} \frac{1}{4}\left[1 + \frac{2\pi^2}{3}\,\tau^2 + O(\tau^2 e^{-1/\tau}\right]  \!\!\!& \text{for $\tau \ll1$}\\
                                                       \tau \left[1 + 1/36\tau^2 + O(1/\tau^4)\right] & \text{for $\tau \gg 1$.}
                               \end{cases}
\label{eq:4.6b}
\ee       
\ese       
We see that the equal-time density-density correlation function, or structure factor, at $T=0$ is a linear function of the wave number, 
and hence long-ranged, even for noninteracting electrons, whereas the static response function at zero wave number
is a constant equal to $\partial n/\partial\mu$. Furthermore, the response function is an analytic function of the
wave number at $q=0$ even in the presence of electron-electron interactions, since the relevant conjugate field,
viz., a change of the chemical potential, does not change the soft-mode spectrum of the system. This illustrates
that (1) long-ranged correlations are common in zero-temperature Fermi systems, and (2) correlation functions
are much more susceptible to this phenomenon than response functions.    

It is important to note that the zero-sound mode in a Fermi liquid does not change the scaling behavior of
the density-density correlation function, it just contributes to the prefactor of the asymptotic linear-in-$q$
behavior. This is because zero sound has the same scaling behavior as the continuum of particle-hole
excitations, viz., $\omega \sim q$. Analogously, the phonon excitations in an interacting Bose system, which
also show $\omega \sim q$ scaling, lead to a linear structure factor. By contrast, a noninteracting Bose gas
behaves very differently since it lacks the particle-hole excitations characteristic of a Fermi gas and their
$\omega \sim q$ scaling.\cite{Landau_Lifshitz_IX_1991}

\paragraph{Number fluctuations, and entanglement entropy}
\label{par:IV.B.1b}

Another manifestation of long-ranged correlations is the entanglement entropy of the fermions, where they
lead to a logarithmically violated area law.\cite{Wolf_2006, Gioev_Klich_2006} The scaling of the
entanglement entropy is closely related to that of particle-number fluctuations, which are easier to calculate
and study.\cite{Klich_2006, Song_Rachel_LeHur_2010, Puspus_Villegas_Paraan_2014, Kwon_Cha_2023}
Consider density fluctuations $\delta n$ in a subvolume $W$ of a large system with volume $V\to\infty$. 
Then the particle-number fluctuation in the subvolume is
\be
\Delta N = \int_W d{\bm x}\,\delta n({\bm x})\ .
\label{eq:4.7}
\ee
and the mean-square fluctuation is
\be
\langle (\Delta N)^2\rangle = \int_W d{\bm x} \int_W d{\bm y}\,\langle\delta n({\bm x})\,\delta n({\bm y})\rangle\ .
\label{eq:4.8}
\ee
If we take $W$ to be a sphere of radius $R$ we obtain, for 3-d systems,
\bse
\label{eqs:4.9}
\be
\langle (\Delta N)^2\rangle = (4\pi)^3 R^3 \int_0^{\kF R} dx\,x^2\,f(x)\,S(q=x/R)
\label{4.9a}
\ee
with
\be
f(x) = \frac{1}{x^6} \left[\sin x - x \cos x\right]^2\ .
\label{eq:4.9b}
\ee
\ese 
and $S(q)$ the equal-time density-density correlation function. At any nonzero temperature we have
$S(q=0) = T\partial n/\partial\mu$, see Eqs.~(\ref{eqs:4.4}) - (\ref{eqs:4.6}), and the mean-square 
number fluctuation scales as the volume of the region $W$, as expected for systems with
short-ranged correlations at finite temperature. However, at $T=0$ we have $S(q\to 0) = (\vF q/4)\partial n/\partial\mu$,
and as a result
\be
\langle (\Delta N)^2\rangle = R^2 8\pi^3 \vF\,\frac{\partial n}{\partial\mu} \left[\ln(\kF R) + \text{const.}\right]\ .
\label{eq:4.10}
\ee
This is an example of the logarithmic violation of the area law in fermion systems. The
entanglement entropy shows the same scaling behavior.\cite{Wolf_2006, Gioev_Klich_2006}
In interacting boson systems the phonons lead to a linear wave-number dependence of
the equal-time density-density correlation function, and hence the analog of the fluctuation
formula (\ref{eq:4.10}) is valid, but the relation to the entanglement entropy is more
complicated.\cite{Laflorencie_2016}

\subsubsection{Origin of the soft modes}
\label{subsubsec:IV.B.2}        

The soft modes causing the long-ranged correlations that are reflected in both correlation functions and
response functions all originate from the soft particle-hole excitations that are reflected in Eqs.~(\ref{eqs:2.5})
and underly, for instance, the scaling behavior of the Lindhard function. Such a continuum of soft modes that
obey a linear scaling of the frequency with the wave number is characteristic of the collisionless regime
in a Fermi liquid, which at low temperatures is realized in a large region of frequency-wave-number space
that excludes only asymptotically low frequencies, and at zero temperature extends all the way to zero
frequency and wave number.\cite{Pines_Nozieres_1989, Belitz_Kirkpatrick_2022, unparticle_footnote} 

In correlation functions these scale-invariant excitations directly lead to long-ranged behavior even in
noninteracting systems, as we
have discussed in Sec.~\ref{subsubsec:IV.B.1}. In the context of response functions the situation is
slightly more complicated. The scale invariant excitations may lead to soft modes of the second kind, such
as Eq.~(\ref{eq:2.5c}) with $\sigma = \sigma'$, which are unaffected by the field conjugate to the relevant 
observable, or to soft modes of the first kind, which acquire a mass in the presence of a field. Only the
latter lead to nonanalyticities in the response functions, and for the nonanalyticities to be realized an
electron-electron-interaction must be present. Importantly, the existence of the soft modes can still be
established by analyzing noninteracting electrons, as we have demonstrated throughout this paper. 
This is a manifestation of the correspondence between the excitations in a Fermi liquid and those
in a Fermi gas that underlies Landau Fermi-liquid theory.

\subsubsection{Nature of the quantum phase transition}
\label{subsubsec:IV.B.3}

As we have discussed, the presence of soft modes of the first kind often renders the 
magnetic phase transition at low temperatures first order. It is interesting to contrast these 
low-temperature transitions of first order with those that are second order. At any nonzero
temperature, the asymptotic critical behavior at a second-order transition is that of the
corresponding classical universality class, and only in a window away from the critical point
do the quantum fluctuations manifest themselves.\cite{Hertz_1976, Sachdev_1999}
By contrast, the nature of the first-order transition induced by the soft modes of the first
kind is inherently quantum mechanical even at nonzero temperature. This is because
the very existence of the first-order transition is due to quantum effects. This point has
been discussed in Ref.~\onlinecite{Belitz_Kirkpatrick_2017}.

\subsubsection{Bandstructure effects}
\label{subsubsec:IV.B.4}

For all of the cases we have discussed, the single-particle energy spectrum, when expanded about the
Fermi energy, is linear in the wave number. This may not be the case in systems with nearly flat, or
dispersionless, bands.\cite{Classen_Betouras_2025} Suppose the band structure is such that the
gradient of the single-particle energy at the Fermi surface vanishes, and the leading wave-number
dependence in the vicinity of the Fermi energy is quadratic. Then the energy, temperature, and
conjugate field all scale quadratically with the wave number. From the scaling arguments in
Sec.~\ref{subsubsec:I.B.1} we see that in this case the scaling of clean systems maps onto that
of disordered systems discussed in Sec.~\ref{subsubsec:I.B.1}. However, the sign of the nonanalyticity
does not change, as the arguments given in the last paragraph of Sec.~\ref{subsubsec:I.B.1} still
apply. As a result, the free-energy functional is given by Eqs.~(\ref{eqs:1.13}), but with $v<0$.
Accordingly, the quantum phase transition is of first order even in clean systems with a nonzero
ordering wave vector, including helimagnets, magnetic smectics, and antiferromagnets.
In the presence of quenched disorder a flat band does not change the behavior compared to a
generic band structure, unless the single-particle energy scales as the wave number to a power
that is higher than quadratic.

\subsubsection{Relation to experiment}
\label{subsubsec:IV.B.5}

The predictions of the soft-mode arguments presented in the present paper are in excellent agreement 
with experiment to the extent that experimental results about the nature of the quantum phase transition are available;
for a review that focuses on ferromagnets based on Landau Fermi liquids see Ref.~\onlinecite{Brando_et_al_2016a}. 
The quantum phase transition problem in ferromagnets based on Dirac Fermi liquids has been
predicted to map onto the Landau case,\cite{Kirkpatrick_Belitz_2019a} but the quantum phase
transition so far has not been studied experimentally. In altermagnets, the nematic part of the
order parameter leads to soft modes and properties of the quantum phase transition that map
onto those of magnetic nematics and ferromagnets, see the discussion in Sec.~\ref{subsec:III.E}.
In considering specific materials one also needs to keep in mind that quantitative considerations 
come into play. For instance, in systems with a nonzero ordering wave number $Q$ the energy scale
$\vF Q$ must be sufficiently large in order to render the quantum phase transition second order, see 
the discussion at the end of Sec.~\ref{subsec:III.B} and the remarks at the end of Sec.~\ref{subsubsec:III.D.1}. 
This is why in helimagnets with a long pitch wavelength, 
such as MnSi or FeGe, the transition is first order as it is in ferromagnets, whereas in systems with a 
large ordering wave number, such as N{\'e}el antiferromagnets, the first-order mechanism is
suppressed. Similarly, whether or not a quantum critical point is realized in non-centrosymmetric 
ferromagnets depends on the size of the spin-orbit interaction, see Sec.~\ref{subsubsec:III.A.4} and the 
discussion in Ref.~\onlinecite{Kirkpatrick_Belitz_2020}. Finally, special features of the band structure
can influence the soft-mode spectrum and its effects on the quantum phase transition, see the
discussion in Sec.~\ref{subsubsec:IV.B.4}. This may be particularly relevant for altermagnets,
as van Hove singularities in flat bands can induce an altermagnetic phase.\cite{Classen_Betouras_2025, Yu_et_al_2025}

\begin{appendix}
\section{The eigenproblem for the helimagnetic Hamiltonian}
\label{app:A}

Here we elaborate on the number of distinct quasiparticle resonances in helimagnets. Consider the Hamiltonian
given in Eq.~(\ref{eq:3.22}) and the associated eigenproblem
\be
\sum_{{\bm k}'} \mathcal{H}_{{\bm k}{\bm k}'}\,{\bm v}_{{\bm k}'}= \lambda\,{\bm v}_{\bm k}
\label{eq:A.1}
\ee
For each value of ${\bm k}$ one finds two pairs of eigenvalues,
\bse
\label{eqs:A.2}
\be
\lambda^{({\bm k},\sigma,+)} = \frac{1}{2}\,\left[\xi_{{\bm k}+{\bm Q}} + \xi_{\bm k} 
    - \sigma\sqrt{\left(\xi_{{\bm k}+{\bm Q}} - \xi_{\bm k}\right)^2 + 4 h^2}\right]
\label{eq:A.2a}
\ee
and
\be
\lambda^{({\bm k},\sigma,-)} = \frac{1}{2}\,\left[\xi_{{\bm k}-{\bm Q}} + \xi_{\bm k} - \sigma\sqrt{\left(\xi_{{\bm k}-{\bm Q}} 
   - \xi_{\bm k}\right)^2 + 4 h^2}\right]
\label{eq:A.2b}
\ee
where $\sigma = \pm$. They are related by
\be
\lambda^{({\bm k}+{\bm Q},\sigma,-)} = \lambda^{({\bm k},\sigma,+)}\ .
\ee
\ese
The eigenvectors corresponding to the eigenvalues $\lambda^{({\bm p},\tau,\alpha)}$ are
\bse
\label{eqs:A.3}
\bea
 {\bm v}_{\bm k}^{({\bm p},\tau,+)} &=& \frac{1}{\left[ (\lambda^{({\bm p},\tau,+)} - \xi_{{\bm p}+{\bm Q}})^2 + h^2\right]^{1/2} }\,
 \nonumber\\
&& \hskip -30pt \times \left[ \delta_{{\bm k}{\bm p}}\, (\lambda^{({\bm p},\tau,+)} - \xi_{{\bm p}+{\bm Q}})\,
    \begin{pmatrix} 1 \\ 0 \end{pmatrix}  -  \delta_{{\bm k},{\bm p}+{\bm Q}}\,h\,\begin{pmatrix} 0 \\ 1 \end{pmatrix}\right]
    \nonumber\\
\label{eq:A.3a}\\
  {\bm v}_{\bm k}^{({\bm p},\tau,-)} &=& \frac{1}{\left[ (\lambda^{({\bm p},\tau,-)} - \xi_{{\bm p}-{\bm Q}})^2 + h^2\right]^{1/2} }\,
  \nonumber\\
&& \hskip -30pt \times \left[ \delta_{{\bm k}{\bm p}}\, (\lambda^{({\bm p},\tau,-)} - \xi_{{\bm p}-{\bm Q}})\,
    \begin{pmatrix} 0 \\ 1 \end{pmatrix}  -  \delta_{{\bm k},{\bm p}-{\bm Q}}\,h\,\begin{pmatrix} 1 \\ 0 \end{pmatrix}\right]
    \nonumber\\
\label{eq:A.3b}    
\eea
Each of these two sets of eigenvectors is linearly independent and spans the space, and the two sets are related by
\be
 {\bm v}_{\bm k}^{({\bm p}+{\bm Q},\tau,-)} = \tau\, {\bm v}_{\bm k}^{({\bm p},\tau,+)}\ .
 \label{eq:A.3c}
\ee
\ese
This shows that each of the sets of eigenvalues in Eqs.~(\ref{eq:A.2a}) and (\ref{eq:A.2b}) represents {\em all}
of the eigenvalues; there is no degeneracy. All of this is consistent with the dimensionality of the space.

Each of the equivalent sets of eigenvectors is orthonormal with respect to both columns and rows,
\bse
\label{eqs:A.4}
 \bea
 \sum_{{\bm k},\sigma} v_{i\sigma}^{({\bm p},\tau,\alpha)} \,v_{{\bm k}\sigma}^{({\bm p}',\tau',\alpha)} = \delta_{{\bm p}{\bm p}'}\,\delta_{\tau\tau'}
 \\
 \sum_{{\bm p},\tau} v_{{\bm k}\sigma}^{({\bm p},\tau,\alpha)} \,v_{{\bm k}'\sigma'}^{({\bm p},\tau,\alpha)} = \delta_{{\bm k}{\bm k}'}\,\delta_{\sigma\sigma'}\ ,
\eea
\ese
which can be ascertained by a direct calculation. Accordingly, the orthogonal matrix $\mathcal{U}$ defined by
\be
\mathcal{U}_{{\bm k}{\bm p}}^{\sigma\tau} := v_{{\bm k}\sigma}^{({\bm p},\tau,+)}
\label{eq:A.5}
\ee
diagonalizes the Hamiltonian,
\be
\left(\mathcal{U}^{\text{T}}\mathcal{H}\,\mathcal{U}\right)_{{\bm p}{\bm p}'}^{\tau\tau'} = \delta_{{\bm p}{\bm p}'}\,\delta_{\tau\tau'}
   \lambda^{({\bm p},\tau,+)}\ ,
\label{eq:A.6}
\ee
and the inverse Hamiltonian is obtained as
\bse
\label{eqs:A.7}
\be
\left(\mathcal{H}^{-1}\right)_{{\bm k}{\bm k}'}^{\sigma\sigma'} = \left(\mathcal{U}\,\mathcal{D}^{-1}\,\mathcal{U}^{\text{T}}\right)_{{\bm k}{\bm k}'}^{\sigma\sigma'} 
\label{eq:A.7a}
\ee
with
\be
\left(\mathcal{D}^{-1}\right)_{{\bm p}{\bm p}'}^{\tau\tau'} = \delta_{{\bm p}{\bm p}'}\,\delta_{\tau\tau'}\,\frac{1}{\lambda^{({\bm p},\tau,+)}}\ .
\label{eq:A.7b}
\ee
\ese
The explicit calculation applied to $i\omega_m\mathbb{1} - \mathcal{H}$ yields the Green function as given in
Eqs.~(\ref{eqs:3.23}). The same procedure with $\mathcal{U}$ defined by $v_{{\bm k}\sigma}^{({\bm p},\tau,-)}$ yields the same result.

The point of this exercise in linear algebra is to demonstrate that the four different quasiparticle resonances in Eqs.~(\ref{eqs:3.24})
are consistent with the the fact that are only two eigenvalues per ${\bm k}$-value: The different parts of the Green function,
and hence the quasiparticle resonances, are given by eigenvalues in different parts of wave-vector space that are separated
by the pitch vector ${\bm Q}$.

\section{The Green function for magnetic smectics and antiferromagnets}
\label{app:B}

In the paramagnetic phase of a magnetic smectic, or a N{\'e}el antiferromagnet, Sec.~\ref{subsec:III.D}, the Green function
is given by
\be
\mathcal{G}_{{\bm k}{\bm p}}(i\omega_m) = \left[i\omega_m \sigma_0 - {\mathcal H}\right]^{-1}_{{\bm k}{\bm p}}\ ,
\label{eq:B.1}
\ee
with the Hamiltonian ${\mathcal H}$ given by Eq.~(\ref{eq:3.35}). Its spin structure is very simple,
\be
{\mathcal G}_{{\bm k}{\bm p}}(i\omega_m) = \sigma_0\,G^{(0)}_{{\bm k}{\bm p}}(i\omega_m) - \sigma_1\,G^{(1)}_{{\bm k}{\bm p}}(i\omega_m)\ .
\label{eq:B.2}
\ee
However, in contrast to the case of a helimagnet, Sec.~\ref{subsec:III.B} and Ref.~\onlinecite{Belitz_Kirkpatrick_Rosch_2006a}, 
it contains terms proportional to $\delta_{{\bm k},{\bm p}+n{\bm Q}}$ for all $n\in\mathbb{Z}$. If we treat the nondiagonal
part of the Hamiltonian as a perturbation we can expand in powers of $h^2$ and make contact with the perturbative
treatment of the eigenproblem in Sec.~\ref{subsec:III.D}. A prominent feature in this expansion is a geometric series
of the form
\begin{widetext}
\bse
\label{eqs:B.3}
\bea
R_{\bm k}(i\omega_m) &=& 1 + \left(\frac{h}{2}\right)^2 g_{\bm k} \left(g_{{\bm k}+{\bm Q}} + g_{{\bm k}-{\bm Q}} \right) 
                                                +  \left(\frac{h}{2}\right)^4 g_{\bm k}^2 \left(g_{{\bm k}+{\bm Q}} + g_{{\bm k}-{\bm Q}}\right)^2 + \ldots
\nonumber\\
&=& \frac{1}{1 - (h/2)^2\, g_{\bm k} \left(g_{{\bm k}+{\bm Q}} + g_{{\bm k}-{\bm Q}}\right)}
\label{eq:B.3a}
\eea
where
\be
g_{\bm k} \equiv g_{\bm k}(i\omega_m) = 1/(i\omega_m - \xi_{\bm k})
\label{eq:B.3b}
\ee
\ese
is the free-fermion Green function. Performing this partial resummation one finds
\bse
\label{eqs:B.4}
\bea
G^{(0)}_{{\bm k}{\bm p}}(i\omega_m) &=& g_{\bm k} \Biggl\{
        \delta_{{\bm k}{\bm p}}\left[1 + O(h^4)\right]
 + \delta_{{\bm k},{{\bm p}}+2{\bm Q}} \left(\frac{h}{2}\right)^2 g_{\bm k}\,  g_{{\bm k}-{\bm Q}} \left[ 1 + O(h^2)\right]
\nonumber\\
&& \hskip 20pt                + \delta_{{\bm k},{{\bm p}}-2{\bm Q}} \left(\frac{h}{2}\right)^2 g_{\bm k}\,  g_{{\bm k}+{\bm Q}} \left[ 1 + O(h^2)\right]           
                                       + \delta_{{\bm k},{{\bm p}}\pm4{\bm Q}} \times O(h^4) + \ldots \Biggr\} R_{\bm k}(i\omega_m)
\label{eq:B.4a}\\
G^{(1)}_{{\bm k}{\bm p}}(i\omega_m) &=& g_{\bm k}\,\frac{h}{2} \Biggl\{
        \delta_{{\bm k},{\bm p}+{\bm Q}} \left[1 + O(h^2)\right] 
      + \delta_{{\bm k},{\bm p}-{\bm Q}} \left[1 + O(h^2)\right]
 + \delta_{{\bm k},{\bm p}\pm3{\bm Q}}\,\times O(h^2)            
+ \ldots \Biggr\} R_{\bm k}(i\omega_m)
\label{eq:B.4b}
\eea
\ese
In this approximation the Green function has poles that are given by the poles of $R_{\bm k}(i\omega)$. From
Eqs.~(\ref{eqs:B.3}) we find a cubic equation for the resonance frequencies,
\be
(i\omega_m - \xi_{\bm k})(i\omega_m - \xi_{{\bm k}+{\bm Q}})(i\omega_m - \xi_{{\bm k}-{\bm Q}})
    - \frac{h^2}{4} (2i\omega_m - \xi_{{\bm k}-{\bm Q}} - \xi_{{\bm k}+{\bm Q}}) = 0\ .
\label{eq:B.5}
\ee
\end{widetext}
This is the same equation that we obtained by truncating the eigenproblem at $O(h^2)$, see Eq.~(\ref{eq:3.41b}). 
This result is consistent with the interpretation of the three energies $E_{\bm k}^{(1,2,3)}$ in Eqs.~(\ref{eqs:3.42})
as single-particle energies.

We finally note that in the limit ${\bm Q}\to 0$ all of the Kronecker deltas in Eqs.~(\ref{eqs:B.4}) turn into
$\delta_{{\bm k}{\bm p}}$. In order to obtain the correct Green function for the ferromagnetic case, Eqs.~(\ref{eqs:2.4}),
one therefore has to keep all terms in the infinite series. 
\end{appendix}


\begin{thebibliography}{128}
\bibitem[FS_()]{FS_splitting_footnote}
It is important to distinguish between `spin-split' Fermi surfaces and
  `field-split' ones, as spin may not be the appropriate quantum number to
  label the split sheets of the Fermi surface.
\bibitem[so_()]{so_footnote}
More generally, the spin-orbit term is of the form
  ${\bm\sigma}\cdot{\bm\Omega}({\bm k})$ with ${\bm\Omega}$ an odd function of
  ${\bm k}$. The soft-mode spectra for all of these Hamiltonians are the same.
\bibitem[cou()]{counting_footnote}
The presence of four branches seems surprising given that the Hamiltonian can
  have only two eigenvalues for each value of the wave vector ${\bm k}$. There
  is no contradiction, see Appendix~\ref{app:A}.
\bibitem[HM_()]{HM_mass_footnote}
This statement is true effectively. A detailed analysis of the soft-mode
  structure shows that the mass is a complicated function of the position on
  the Fermi surface.
\end{thebibliography}

\end{document}